\documentclass{aa}  
\pdfoutput=1
\usepackage{graphicx}
\usepackage{txfonts}
\usepackage{longtable} 
\usepackage{natbib}
\bibpunct{(}{)}{;}{a}{}{,} 
\begin{document}

   \title{Complex organics in IRAS\,4A revisited with ALMA and PdBI: Striking contrast between two neighbouring protostellar cores}
   \titlerunning{Complex organics in IRAS\,4A revisited with ALMA and PdBI}


   \author{A. L\'opez-Sepulcre
          \inst{1,2}
          \and
          N. Sakai
          \inst{3}
          \and
          R. Neri
          \inst{1}
          \and
          M. Imai
          \inst{2}
          \and
          Y. Oya
          \inst{2}
          \and
          C. Ceccarelli
          \inst{4,5}
          \and
          A.E. Higuchi
          \inst{3}
          \and
          Y. Aikawa
          \inst{6}
          \and
          S. Bottinelli
          \inst{7,8}
          \and
          E. Caux
          \inst{7,8}
          \and
          T. Hirota
          \inst{9}
          \and
          C. Kahane
          \inst{4,5}
          \and
          B. Lefloch
          \inst{4,5}
          \and
          C. Vastel
          \inst{7,8}
          \and
          Y. Watanabe
          \inst{2}
          \and
          S. Yamamoto
          \inst{2}
          \fnmsep}

   \institute{   Institut de Radioastronomie Millim\'etrique (IRAM), 300 rue de la Piscine, 38406 Saint-Martin-d'H\`eres, France\\
         \email{lopez@iram.fr}
         \and
   Department of Physics, The University of Tokyo, Bunkyo-ku, Tokyo 113-0033, Japan
         \and
   The Institute of Physical and Chemical Research (RIKEN), 2-1, Hirosawa, Wako-shi, Saitama 351-0198, Japan
         \and
   Univ. Grenoble Alpes, IPAG, F-38000 Grenoble, France
         \and
   CNRS, IPAG, F-38000 Grenoble, France
         \and
   Center for Computational Science, University of Tsukuba, Tsukuba, Ibaraki 305-8577, Japan
         \and
   Universit\'e de Toulouse, UPS-OMP, IRAP, Toulouse, France
        \and
   CNRS, IRAP, 9 Av. Colonel Roche, BP 44346, F-31028 Toulouse Cedex 4, France
           \and
   National Astronomical Observatory of Japan, Osawa, Mitaka, Tokyo 181-8588, Japan
                }

   \date{Received ; accepted }

  \abstract
   {Hot corinos are extremely rich in complex organic molecules (COMs). Accurate abundance measurements of COMs in such objects are crucial to constrain astrochemical models. In the particular case of close binary systems this can only be achieved through high angular resolution imaging.}
   {We aim to perform an interferometric study of multiple COMs in NGC1333\,IRAS\,4A, which is a protostellar binary hosting hot corino activity, at an angular resolution that is sufficient to distinguish easily the emission from the two cores\ separated by 1.8$''$.}
   {We used the Atacama Large (sub-)Millimeter Array (ALMA) in its 1.2\,mm band and the IRAM Plateau de Bure Interferometer (PdBI) at 2.7\,mm to image, with an angular resolution of 0.5$''$ (120\,au) and 1$''$ (235\,au), respectively, the emission from 11 different organic molecules in IRAS\,4A. This allowed us to clearly disentangle A1 and A2,  the two protostellar cores. For the first time, we were able to derive the column densities and fractional abundances simultaneously for the two objects, allowing us to analyse the chemical differences between them.}
   {Molecular emission from organic molecules is concentrated exclusively in A2, while A1 appears completely devoid of COMs or even simpler organic molecules, such as HNCO, even though A1 is the strongest continuum emitter. The protostellar core A2 displays typical hot corino abundances and its deconvolved size is 70\,au. In contrast, the upper limits we placed on COM abundances for A1 are extremely low, lying about one order of magnitude below prestellar values. The difference in the amount of COMs present in A1 and A2 ranges between one and two orders of magnitude. Our results suggest that the optical depth of dust emission at these wavelengths is unlikely to be sufficiently high to completely hide a hot corino in A1 similar in size to that in A2. Thus, the significant contrast in molecular richness found between the two sources is most probably real. We estimate that the size of a hypothetical hot corino in A1 should be less than 12\,au.}
   {Our results favour a scenario in which the protostar in A2 is either more massive and/or subject to a higher accretion rate than A1, as a result of inhomogeneous fragmentation of the parental molecular clump. This naturally explains the smaller current envelope mass in A2 with respect to A1 along with its molecular richness. The extremely low abundances of organic molecules in A1 with respect to those in A2 demonstrate that the dense inner regions of a young protostellar core lacking hot corino activity may be poorer in COMs than the outer protostellar envelope.}


   \maketitle

\section{Introduction}\label{intro}

Life on Earth is based on organic chemistry, which very likely had its origin at the very first stages of solar system formation. Indeed, several organic molecules have been detected in cold, solar-mass prestellar cores \citep{oberg10,bacmann12,cernicharo12,vastel14,js16}, indicating that they can be synthesised prior to the onset of star formation. Such molecules include, for example acetaldehyde (CH$_3$CHO), dimethyl ether (CH$_3$OCH$_3$), and methyl formate (HCOOCH$_3$), and they are known by the astrochemistry community as complex organic molecules (COMs).

Complex organic molecules were first discovered in hot molecular cores associated with high-mass star-forming regions by \citet{fourikis74}, \citet{brown75}, and \citet{snyder74}. Almost three decades later, the discovery of hot corinos followed. These are compact ($<\,100$\,au), dense ($> 10^{7}$\,cm$^{-3}$), and hot ($> 100$\,K) regions in the immediate vicinity of low-mass protostars where saturated COMs are abundant \citep[e.g.][]{cazaux03,bottinelli04b,jorgensen05,sakai06,oberg11,codella16,imai16,desimone17}.

Both dust grain surface chemistry \citep[e.g.][]{garrod06,garrod08} and gas-phase chemistry \citep[e.g.][]{rodgers01,balucani15,barone15,er16} have been proposed to explain the extreme molecular richness present in hot corinos. Constraining astrochemical models and their chemical networks requires accurate measurements of COMs abundances in these objects, a task that is often hindered by their small sizes ($\leq 100$\,au). Indeed, COMs can also be present in the more external regions of protostellar envelopes \citep[e.g.][]{jaber14} and in shocks produced by fast jets and molecular outflows \citep[e.g.][]{arce08,sugimura11}. The situation becomes even more complicated if the source is a binary, as is the case of I16293 and NGC\,1333~IRAS\,4A. In order to isolate the emission originating from hot corinos, it is important to make use of (sub-)arcsecond interferometric imaging. While various interferometric studies of hot corinos have been carried out so far \citep[e.g.][]{bottinelli04a,kuan04,bisschop08,jorgensen11,maury14,codella16,oya16,jorgensen16}, most of them concentrate on a rather limited sample of COMs.

In this work, we present interferometric observations of multiple COMs in the protostellar binary NGC\,1333~IRAS\,4A (hereafter IRAS\,4A). IRAS\,4A is located in the Perseus star formation complex at a distance of 235\,pc \citep{hirota08}. The total envelope mass and luminosity of this object are estimated to be 5.6\,M$_\odot$ and 9.1\,L$_\odot$, respectively \citep{kristensen12,karska13}. While the two cores, named A1 and A2, originate from the same parent cloud and drive similarly powerful molecular outflows, oriented roughly in a south-north direction \citep{santangelo15}, only A2 harbours a hot corino \citep{persson12}. \citet{taquet15} performed a comprehensive study of numerous COMs in IRAS\,4A with PdBI. However, the angular resolution of their observations ($\sim 2''$) was not sufficient to clearly disentangle A1 from A2 and thus provide accurate molecular abundances for the latter. We here improve the angular resolution by a factor 4 to 16 in area. This allows us, for the first time, to spatially separate the emission of a relatively large number of COMs in A2 and to provide upper limits on the COMs abundances of A1.

The present paper is structured as follows. Section\,\ref{obs} describes our observations and data reduction. In Sect\,\ref{results} we present our continuum and molecular maps, as well as molecular spectra, and we derive molecular abundances. In Sect\,\ref{discussion} we discuss our results, with a particular emphasis on the large difference in molecular emission between A1 and A2. Finally, our main conclusions are summarised in Sect.\,\ref{conclusions}.

\section{Observations}\label{obs}

\subsection{ALMA 1.2\,mm observations}

The Atacama Large (sub-)Millimeter Array (ALMA) was used on 14 June 2014, in its Cycle\,2 operations, to observe IRAS\,4A in Band\,6 ($\sim 250$\,GHz). A total of 35 antennas of the 12 m array were used with a minimum and maximum baseline length of 18\,m and 784\,m, respectively. The phase centre of the observation was R.A.(J2000)\,$=$\,03$^\mathrm{d}$29$^\mathrm{m}$10.51$^\mathrm{s}$, Dec.(J2000)\,$=$+31$^\circ$13$'$31.3$''$ and the total integration time on-source was 26\,min. We employed 16 spectral windows between 244 and 264\,GHz, whose bandwidth and spectral resolution are 58.6\,MHz (70\,km\,s$^{-1}$) and 122 kHz (0.14\,km\,s$^{-1}$), respectively. The system temperature ranged between 75 and 105\,K. The bandpass, flux, and phase calibrators used were J0237+2848, J0238+166, and J0319+4130, respectively. The flux calibration error is estimated to be 10\%.

We reduced the data with the Common Astronomy Software Applications package (CASA) and subsequently employed GILDAS\footnote{http://www.iram.fr/IRAMFR/GILDAS/} for image and spectral analysis. We obtained a continuum image by averaging line-free channels from all the spectral windows. We then generated spectral cubes by subtracting the continuum emission directly in the visibility plane. We CLEANed the continuum image and spectral cubes using Briggs weighting with the robustness parameter set to 0.5. We performed phase self-calibration of the continuum data and applied the solutions to both the continuum and molecular lines. This greatly improved the signal-to-noise ratio, particularly in the continuum map. The resulting synthesised beam for the continuum image is $0.66'' \times 0.35''$ (P.A.\,$=$\,$-29^\circ$). The spectral images have similar beam sizes (see Sect.\,\ref{mol}). At the frequency of the observations, the primary beam size is 23$''$ and the largest scale recovered by the array is 3.3$''$.

\subsection{IRAM PdBI 2.7\,mm observations}

The PdBI was used on 29 January 2005 in its A configuration to observe IRAS\,4A at 2.7\,mm. The baselines ranged from 32 to 400\,m. The phase centre of the observations was R.A.(J2000)\,$=$\,03$^\mathrm{d}$29$^\mathrm{m}$10.30$^\mathrm{s}$, Dec.(J2000)\,$=$+31$^\circ$13$'$31.0$''$. Continuum emission as well as five CH$_3$CN transitions at 110.4\,GHz were observed. The latter were covered with two correlator units, each with a bandwidth of 40\,MHz and a spectral resolution of 78\,kHz or 0.2\,km\,s$^{-1}$. The system temperature ranged typically between 150 and 250\,K. The bandpass and flux calibrators used were 3C454.3 and MWC349, respectively. 0234+285 and 3C84 were observed as phase/amplitude calibrators every 20 minutes throughout the entire run. The flux calibration uncertainty is $\sim$10\%.

The data were reduced using the packages CLIC and MAPPING of the GILDAS software collection. We produced a continuum image by averaging line-free channels and self-calibrating the data. The final cleaned continuum map was obtained using robust weighting parameter equal to 0.1. The self-calibration solutions were then applied to the spectral cube, which was subsequently cleaned using natural weighting to maximise sensitivity. The resulting synthesised beam for the continuum image is $1.17'' \times 1.01''$ (P.A.\,$=$\,$35^\circ$) and that of the spectral cube is $1.35'' \times 1.04''$ (P.A.\,$=$\,$-168^\circ$).

\section{Results and analysis}\label{results}

\subsection{Dust continuum emission}\label{cont}

\begin{figure*}
\centering
\begin{tabular}{lr}
\includegraphics[scale=0.38]{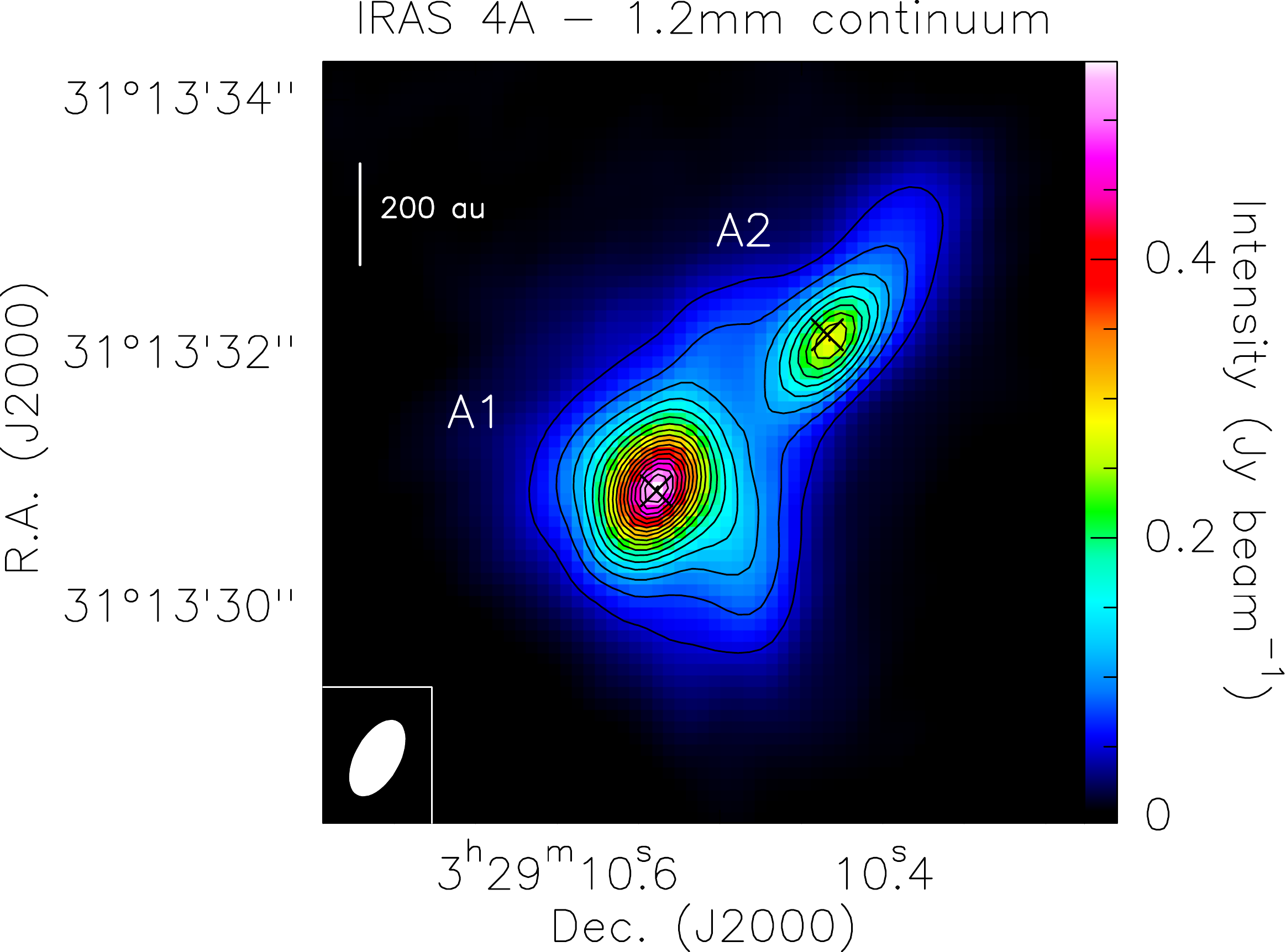} & \includegraphics[scale=0.38]{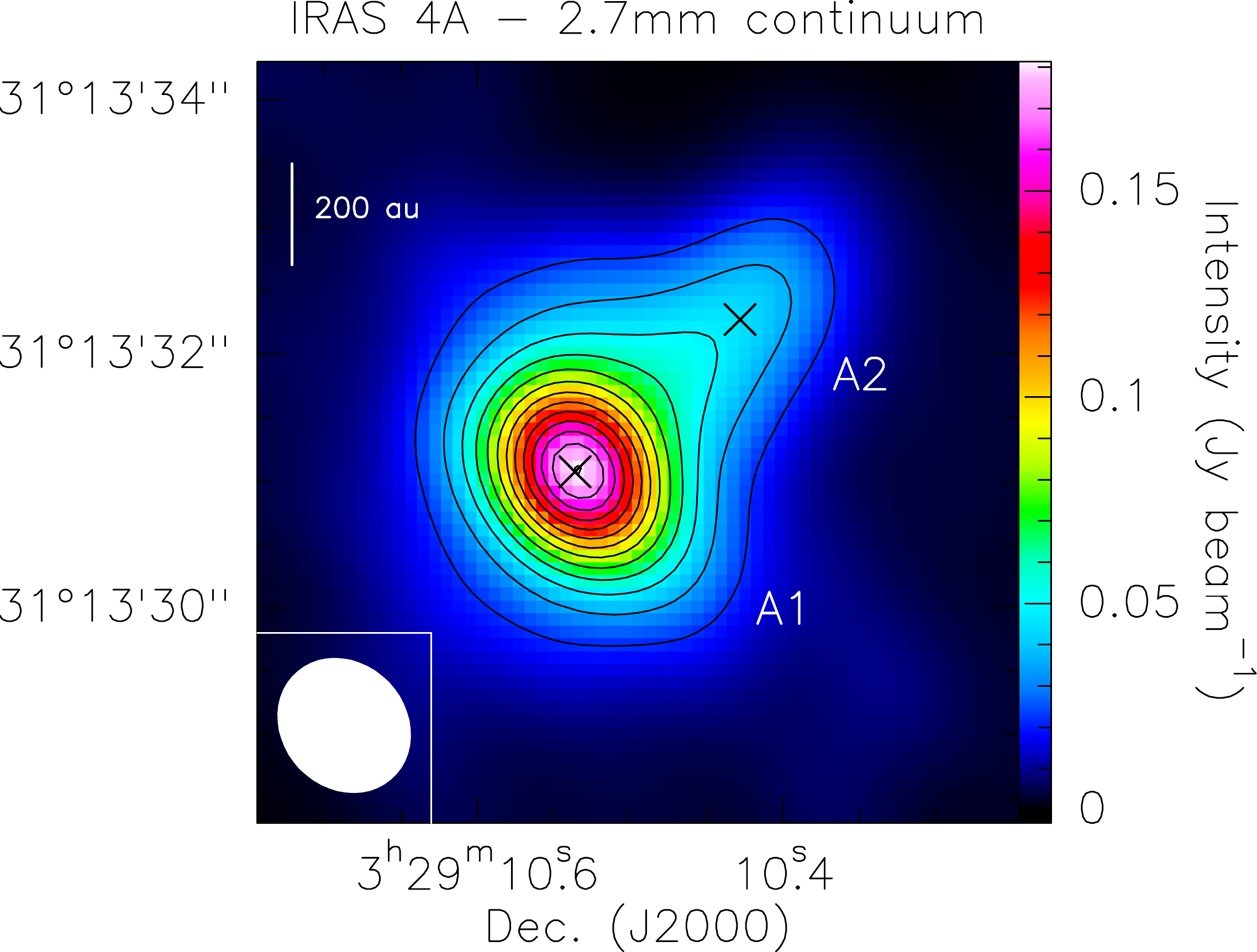}\\
\end{tabular}
\caption{Dust continuum emission maps of IRAS\,4A at 1.2\,mm (\textit{left}) and 2.7\,mm (\textit{right}). For the two maps, contours start at 50$\sigma$ and increase by steps of 30$\sigma$, with $\sigma = 0.96$\,mJy\,beam$^{-1}$ and 0.44\,mJy\,beam$^{-1}$ at 1.2 and 2.7\,mm, respectively. The two protostellar cores A1 and A2 are labelled and their coordinates, obtained from 2D Gaussian fits (see Table\,\ref{tmass}), are indicated with black crosses. The synthesised beams are depicted in white in the lower left corner of each panel.}
\label{fcont}
\end{figure*}

Figure\,\ref{fcont} presents the 1.2 and 2.7\,millimetre continuum maps of IRAS\,4A. The two protostellar cores, A1 (south-east) and A2 (north-west), are clearly separated at the sub-arcsecond resolution of our ALMA 1.2\,mm observations, while they are marginally resolved in the 2.7\,mm image. The angular separation between the two is 1.8$''$, which translates into a spatial separation of 420\,au at a distance of 235\,pc. This is in good agreement with previous interferometric studies \citep[e.g.][]{lay95,looney00}.

The root mean square (RMS) noise levels of the ALMA and PdBI continuum emission maps are 0.96\,mJy\,beam$^{-1}$  and 0.44\,mJy\,beam$^{-1}$, respectively, and the total fluxes within the 5$\sigma$ contour levels are 3.88\,$\pm$\,0.39\,Jy and 0.56\,$\pm$\,0.06 Jy, respectively. The uncertainties in the flux measurements include both the RMS and the amplitude calibration errors and are largely dominated by the latter. The deconvolved sizes and fluxes of A1 and A2, obtained from 2D Gaussian fits in the visibility plane, are listed in Table\,\ref{tmass}. The resulting fluxes are consistent within 30\% with those measured by \citet{persson12} and \citet{santangelo15}. Using the deconvolved fluxes, we can provide an estimate of the mass of the protostellar cores under the assumption of isothermality and optically thin dust emission from the following equation:

\begin{equation}
M_\mathrm{gas} = \frac{S_\nu d^2}{\kappa_\nu B_\nu(T_\mathrm{d}) R_\mathrm{d}},
\label{emass}
\end{equation}
where $S_\nu$ is the flux measured at the observed frequencies (250\,GHz and 110\,GHz), $d = 235$\,pc the distance to IRAS\,4A, $T_\mathrm{d}$ the dust temperature, $B_\nu (T_\mathrm{d})$ the Planck black-body function for a temperature $T_\mathrm{d}$, and $R_\mathrm{d}$ the dust-to-gas mass ratio, adopted to be 0.01.\ The value $\kappa_\nu$ is the frequency-dependent dust mass opacity coefficient, which we assume to be 1.3 and 0.30\,cm$^2$\,g$^{-1}$ at wavelengths of 1.2 and 2.7\,mm, respectively. These values of  $\kappa_\nu$ are obtained by simple linear interpolation (1.2\,mm) and extrapolation (2.7\,mm), in logarithmic scale, of the tabulated values given in \citet{ossenkopf94}, for thin icy mantles and a gas density of 10$^8$\,cm$^{-3}$. If thick icy grain mantles are considered instead, which might be the case in A1 (see Sect.\,\ref{da1}), the resulting masses increase by 9\%, which is a small factor compared to the much larger uncertainties introduced by the unknown dust temperature. 

For the purpose of deriving molecular abundances in Sect.\,\ref{deriv}, we computed the beam-averaged H$_2$ column densities at the peak coordinates of A1 and A2, as follows:

\begin{equation}
N_\mathrm{H2} = \frac{S_{\nu\mathrm{,peak}}}{\kappa_\nu \mu m_\mathrm{H} B_\nu(T_\mathrm{d}) R_\mathrm{d} \Omega_\mathrm{beam}},
\label{ecol}
\end{equation}
where $S_\nu$ is the peak flux density in each source, $m_\mathrm{H}$ is the mass of the hydrogen atom, $\mu = 2.33$ is the mean molecular mass in units of hydrogen atom masses, and $\Omega_\mathrm{beam}$ is the solid angle subtended by the synthesised beam. The results thus obtained for the mass and column densities of A1 and A2 are summarised in Table\,\ref{tmass}, and they correspond to a range of dust temperatures between $T_\mathrm{d} = 50$ and 200\,K (highest and lowest mass and column density values, respectively). We are aware that the assumption of isothermality is a simplification, as large temperature gradients are to be expected at smaller scales than those probed by the beam of our observations. However, without a small-scale model of the physical structure of IRAS\,4A, adopting a reasonable range of beam-averaged dust temperatures is the best we can currently do with our data at this stage.

\begin{table}[!ht]
\centering
\caption{Dust continuum parameters of IRAS\,4\,A1 and A2$^\mathrm{a}$}
\begin{tabular}{lcc}
\hline
\multicolumn{3}{c}{ALMA 1.2\,mm}\\
\hline
 & A1 & A2 \\
\cline{2 - 3}
R.A. (J2000) & 03$^\mathrm{h}$29$^\mathrm{m}$10.539$\mathrm{s}$ & 03$^\mathrm{d}$29$^\mathrm{m}$10.434$\mathrm{s}$\\
Dec. (J2000) & +31$^\circ$13$'$30.92$''$ & +31$^\circ$13$'$32.15$''$\\
Size ($''$)$^\mathrm{b}$ & $1.0 \times 0.8$ & $1.4 \times 0.8$\\
P.A. ($^\circ$)$^\mathrm{b}$ & 36 & 137\\
$S_{\nu}$ (Jy)$^\mathrm{b}$ & 1.99 $\pm$ 0.20 & 1.15 $\pm$ 0.12\\
$S_{\nu\mathrm{,peak}}$ (mJy\,beam$^{-1}$) & 542 $\pm$ 55 & 279 $\pm$ 28\\
$M_\mathrm{gas}$ (M$_\odot$) & 0.09 -- 0.51 & 0.06 -- 0.29\\
$N_\mathrm{H2,peak}$ ($10^{24}$\,cm$^{-2}$)$^\mathrm{c}$ & 5.7 -- 31 & 2.9 -- 16\\
\hline
\multicolumn{3}{c}{PdBI 2.7\,mm}\\
\hline
 & A1 & A2 \\
\cline{2 - 3}
R.A. (J2000) & 03$^\mathrm{h}$29$^\mathrm{m}$10.536$\mathrm{s}$ & 03$^\mathrm{d}$29$^\mathrm{m}$10.428$\mathrm{s}$\\
Dec. (J2000) & +31$^\circ$13$'$31.07$''$ & +31$^\circ$13$'$32.27$''$\\
Size ($''$)$^\mathrm{b}$ & 1.3 $\times$ 0.9 & 1.8 $\times$ 1.3\\
P.A. ($^\circ$)$^\mathrm{b}$ & 29 & 137\\
$S_{\nu}$ (Jy)$^\mathrm{b}$ & 0.345 $\pm$ 0.20 & 0.162 $\pm$ 0.12\\
$S_{\nu\mathrm{,peak}}$ (mJy\,beam$^{-1}$) & 167 $\pm$ 17 & 46 $\pm$ 5\\
$M_\mathrm{gas}$ (M$_\odot$) & 0.37 -- 1.9 & 0.17 -- 0.88\\
$N_\mathrm{H2,peak}$ ($10^{24}$\,cm$^{-2}$)$^\mathrm{c}$ & 7.7 -- 40 & 2.1 -- 11\\
\hline
\end{tabular}
\\
\raggedright $^\mathrm{a}$ Masses and H$_2$ column densities derived for a dust temperature range of 50 -- 200\,K.\\
$^\mathrm{b}$ Deconvolved source sizes, coordinates, and fluxes from 2D Gaussian fits.\\
$^\mathrm{c}$ Beam-averaged H$_2$ column densities at each peak position.
\label{tmass}
\end{table}

The H$_2$ column densities obtained are of the order of a few times $10^{24}$\,cm$^{-2}$. Assuming the cores have the same size along the line of sight as they do on the plane of the sky, this translates into a density of $\sim 10^8$\,cm$^{-3}$, which is consistent with our choice of $\kappa_\nu$. At 1.2\,mm, the dust becomes optically thick at $N_\mathrm{H2} \geq 2 \times 10^{25}$\,cm$^{-2}$ for the adopted dust opacity coefficient and dust-to-gas mass ratio. As a result, the assumption of optically thin dust may not hold in A1 at this wavelength, especially if the dust temperature lies close to the low limit of the adopted range. Indeed, the upper column density estimate for A1 translates into $\tau \approx 1.6$. The masses and column densities derived for A1 should therefore be regarded as lower limits. For A2, the assumption of optically thin dust is reasonable at the mm wavelengths considered in this study, although it might be marginally optically thick at 1.2\,mm ($\tau \approx 0.8$) if the dust temperature is sufficiently low. A more detailed discussion on this issue is presented in Sect.\,\ref{da1}.

The 2.7\,mm mass estimates are larger by a factor $\sim$\,3 than those computed from the 1.2\,mm data. Different factors may cause such discrepancies, such as the larger deconvolved sizes obtained at 2.7\,mm, the uncertainty on $\kappa_\nu$ at this wavelength, and possibly a larger portion of filtered out flux at 1.2\,mm. As for the column densities, the discrepancies are smaller, but one must bear in mind that the synthesised beams at 1.2 and 2.7\,mm have different areas. For the sake of comparison, we measured the column densities averaged over a circle with a diameter of 1.2$''$ centred around the peak coordinates of A1 and A2. Adopting $T_\mathrm{d} = 100$\,K we obtained at 1.2 and 2.7\,mm, $N_\mathrm{H2} = 4.5 \times 10^{24}$\,cm$^{-2}$ and $N_\mathrm{H2} = 9.8 \times 10^{24}$\,cm$^{-2}$ for A1, and $N_\mathrm{H2} = 1.9 \times 10^{24}$\,cm$^{-2}$ and $N_\mathrm{H2} = 2.7 \times 10^{24}$\,cm$^{-2}$ for A2, respectively. Thus, at 2.7\,mm, the column density of A1 is a factor 2 larger than that at 1.2\,mm, while for A2 it is a factor 1.4 larger. Again, uncertainties in $\kappa_\nu$ and differences in the amount of filtered out flux between the two wavelengths may contribute to the discrepancy. However, the fact that this difference is larger for A1 than for A2 might be an indication of optically thick dust at 1.2\,mm in A1, which results in attenuation of the emission at this wavelength and therefore a lower value of the column density.

\subsection{Molecular line emission}\label{mol}

We employed the Cologne Database for Molecular Spectroscopy \citep{muller01,muller05} and the Jet Propulsion Laboratory spectroscopic database \citep{pickett98} to identify the observed spectral lines from COMs and simpler organic molecules, such as HNCO. In total, we detect 23 well-isolated lines with $S/N > 4$ for a channel spacing of 0.2\,km\,s$^{-1}$, belonging to 11 different molecular species. These include two torsionally excited transitions of HCOOCH$_3$. We additionally detect lines from the following species and transitions: CH$_3$CHO($14_{1,14}-13_{1,13}$)A, C$_2$H$_5$CN($29_{5,25}-28_{5,24}$), (CH$_3$)$_2$CO($14_{11,3}-13_{10,4}$)AE, NH$_2$CHO($12_{0,12}-11_{0,11}$)v$_{12}$\,$=$\,1. All of these appear blended with other molecular lines and are thus not considered in our analysis. The list of the 17 identified lines is presented in Table\,\ref{tlines}, together with their respective Gaussian fit parameters. For HCOOH, CH$_2$OHCHO, C$_2$H$_5$OH, and HNCO, only one transition is clearly detected. Despite this, we stress that the detection of these species is solid for two reasons. First, they have been detected in other observational studies of the same object \citep[e.g.][]{bottinelli04b,jorgensen11,taquet15,yo15}. Second, the ALMA observations are part of a larger project that includes other objects observed with the same frequency set-up. As a result, many of the lines we detect in IRAS\,4A are also detected in other sources, in particular in B335 \citep{imai16}, which lends further validity to our line identification. We thus conclude that all the detections in the present study are robust.

\begin{table*}
\centering
\caption{Complex organic molelcule lines detected in IRAS\,4A2 and Gaussian fit results}
\begin{tabular}{llcccccccl}
\hline
Molecule & Transition & $\nu_0$ & $E_\mathrm{up}$ & $S \mu^2$$^\mathrm{a}$ & Flux & $V_\mathrm{lsr}$ & $\Delta V$ & $T_\mathrm{peak}$ & Note$^\mathrm{b}$\\
  & & (GHz) & (K) & (D$^2$) & (Jy\,km\,s$^{-1}$) & (km\,s$^{-1}$) & (km\,s$^{-1}$) & (mJy\,beam$^{-1}$) & \\
\hline
CH$_3$CN & $6_5-5_5$ & 110.33035 & 197.1 & 38.9 & 0.07 $\pm$ 0.02 & 5.6 $\pm$ 0.3 & 1.5 $\pm$ 0.6 & 43 $\pm$ 28 & \\
CH$_3$CN & $6_4-5_4$ & 110.34947 & 132.8 & 70.7 & 0.19 $\pm$ 0.02 & 6.4 $\pm$ 0.2 & 2.8 $\pm$ 0.5 & 64 $\pm$ 17 & \\
CH$_3$CN & $6_3-5_3$ & 110.36435 & 82.8 & 191 & 0.32 $\pm$ 0.02 & 6.9 $\pm$ 0.1 & 3.4 $\pm$ 0.3 & 91 $\pm$ 17 & \\
CH$_3$CN & $6_2-5_2$ & 110.37499 & 47.1 & 113 & 0.30 $\pm$ 0.02 & 6.9 $\pm$ 0.1 & 3.0 $\pm$ 0.3 & 94 $\pm$ 17 & \\
CH$_3$CN & $6_1-5_1$ & 110.38137 & 25.7 & 123 & 0.26 $\pm$ 0.02 & 6.8 $\pm$ 0.1 & 2.5 $\pm$ 0.3 & 101 $\pm$ 17 & \\
CH$_3$CN & $6_0-5_0$ & 110.38350 & 18.5 & 127 & 0.28 $\pm$ 0.02 & 7.0 $\pm$ 0.1 & 2.9 $\pm$ 0.4 & 89 $\pm$ 17 & \\
\hline
t-HCOOH & $12_{0,12}-11_{0,11}$ & 262.10348 & 82.8 & 24.2 & $0.42 \pm 0.02$ & $6.9 \pm 0.1$ & $3.8 \pm 0.2$ & $100 \pm 10$ & \\ 
CH$_3$CHO & $14_{1,14}-13_{1,13}$\,E & 260.53040 & 96.4 & 81.6 & $0.61 \pm 0.08$ & $6.7 \pm 0.1$ & $2.1 \pm 0.4$ & $270 \pm 70$ & \\ 
HCOOCH$_3$ & $20_{9,12}-19_{9,11}$\, E\,v$_\mathrm{t}$\,$=$\,1 & 245.54388 & 364.6 & 42.5 & $0.15 \pm 0.02$ & $6.9 \pm 0.1$ & $2.2 \pm 0.3$ & $62 \pm 16$ & \\ 
HCOOCH$_3$ & $20_{16,4}-19_{16,3}$\,A & 245.57471 & 293.6 & 38.3 & $0.14 \pm 0.03$ & $6.9 \pm 0.2$ & $1.8 \pm 0.5$ & $72 \pm 27$ & 2T\\ 
HCOOCH$_3$ & $20_{3,17}-19_{3,16}$\,A\,v$_\mathrm{t}$\,$=$\,1 & 248.71584 & 320.9 & 51.0 & $0.13 \pm 0.01$ & $7.0 \pm 0.1$ & $1.6 \pm 0.1$ & $75 \pm 8$ & \\ 
HCOOCH$_3$ & $20_{5,16}-19_{5,15}$\,E & 249.03100 & 141.6 & 49.7 & $0.37 \pm 0.01$ & $6.9 \pm 0.1$ & $1.9 \pm 0.1$ & $180 \pm 10$ & \\ 
HCOOCH$_3$ & $20_{5,16}-19_{5,15}$\,A & 249.04743 & 141.6 & 49.7 & $0.44 \pm 0.02$ & $6.9 \pm 0.1$ & $2.2 \pm 0.1$ & $190 \pm 20$ & \\ 
CH$_3$OCH$_3$ & $13_{5,8}-13_{4,9}$\,AE & 262.38922 & 118.0 & 64.8 & $0.11 \pm 0.01$ & $6.9 \pm 0.1$ & $1.6 \pm 0.2$ & $63 \pm 15$ & \\ 
CH$_3$OCH$_3$$^\mathrm{c}$ & $13_{5,8}-13_{4,9}$ & 262.39325 & 118.0 & 283 & --- & --- & --- & --- & 3T\\ 
CH$_3$OCH$_3$ & $30_{4,26}-30_{3,27}$ & 264.21161 & 446.9 & 856 & $0.12 \pm 0.01$ & $7.1 \pm 0.1$ & $2.5 \pm 0.2$ & $46 \pm 8$ & 4T\\ 
(CH$_3$)$_2$CO & $22_{3,19}-21_{4,18}$\,AE & 246.40427 & 149.6 & 2302 & $0.08 \pm 0.01$ & $7.1 \pm 0.1$ & $1.8 \pm 0.3$ & $43 \pm 12$ & 4T\\ 
(CH$_3$)$_2$CO & $24_{1,23}-23_{2,22}$\,AA & 248.70330 & 156.2 & 3048 & $0.12 \pm 0.01$ & $7.1 \pm 0.1$ & $2.2 \pm 0.2$ & $52 \pm 8$ & 2T\\ 
cis-CH$_2$OHCHO & $25_{2,24}-24_{1,23}$ & 262.05678 & 170.8 & 111 & $0.22 \pm 0.01$ & $6.8 \pm 0.1$ & $2.5 \pm 0.1$ & $83 \pm 8$ & \\ 
C$_2$H$_5$OH & 14$_{3,11}-13_{3,10}$ & 246.41476 & 155.7 & 21.4 & $0.20 \pm 0.03$ & $6.8 \pm 0.1$ & $2.1 \pm 0.3$ & $92 \pm 28$ & \\
C$_2$H$_5$CN & $29_{2,27}-28_{2,26}$ & 263.81079 & 194.71 & 427 & $0.13 \pm 0.01$ & $6.9 \pm 0.1$ & $1.7 \pm 0.2$ & $68 \pm 12$ & \\ 
NH$_2$CHO & $12_{0,12}-11_{0,11}$ & 247.39072 & 78.1 & 157 & $0.40 \pm 0.01$ & $7.1 \pm 0.1$ & $4.2 \pm 0.1$ & $89 \pm 6$ & \\ 
HNCO & $12_{0,12}-11_{0,11}$ & 263.74863 & 82.3 & 30.0 & $0.82 \pm 0.03$ & $6.9 \pm 0.1$ & $3.8 \pm 0.2$ & $200 \pm 30$ & \\ 
\hline
\end{tabular}
\raggedright $^\mathrm{a}$ For those lines with several blended transitions of the same species at (almost) the same rest frequencies, this is the sum of $S \mu^2$ of all the transitions involved.\\
$^\mathrm{b}$ \#T: blended with \# transitions of the same molecule.\\
$^\mathrm{c}$ Gaussian fit not performed owing to blending of different CH$_3$OCH$_3$ transitions (see Fig.\,\ref{fspt}).
\label{tlines}
\end{table*}

\begin{figure*}[!ht]
\centering
\includegraphics[scale=0.9]{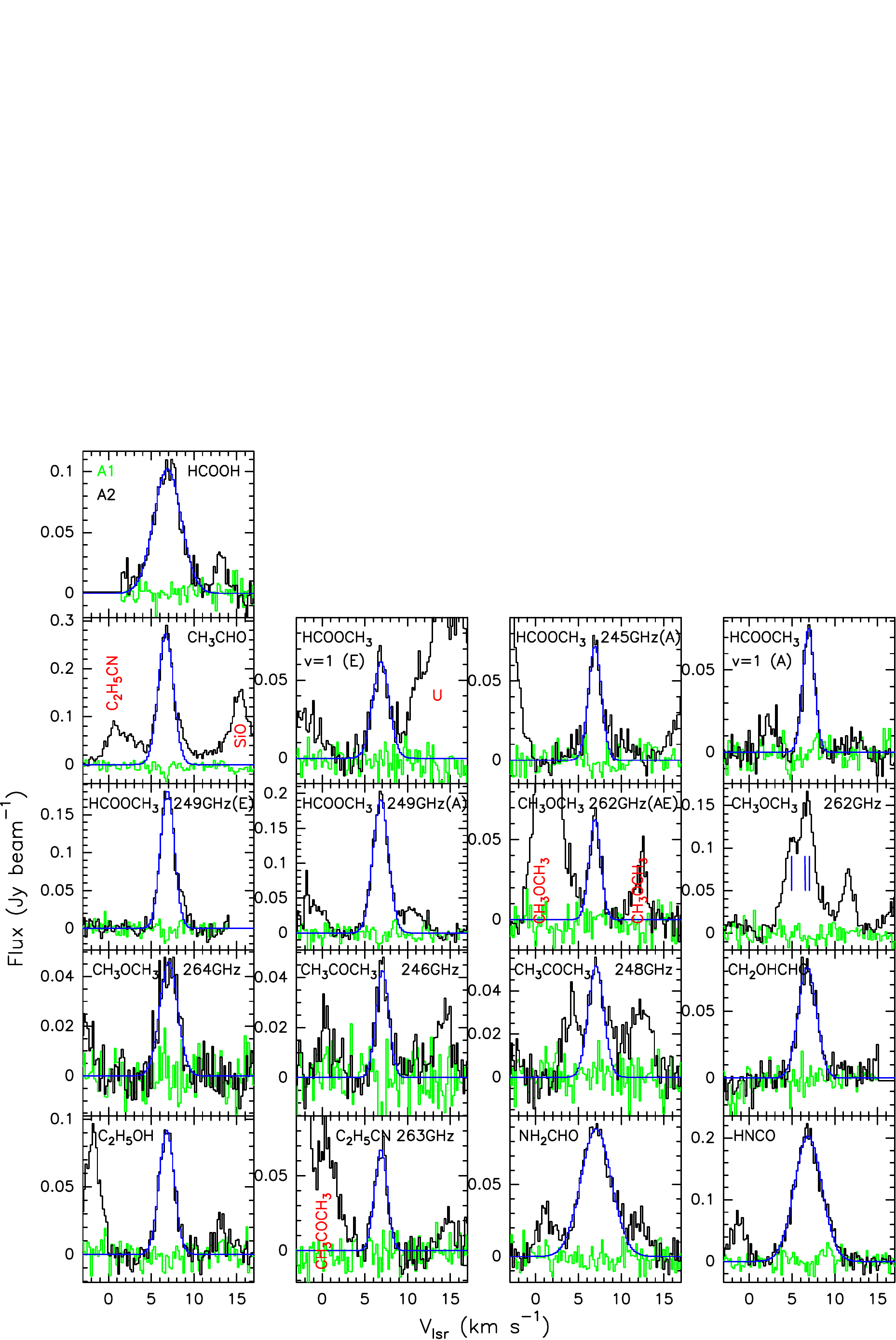}
\caption{Spectra of the 17 lines from organic molecules identified in IRAS\,4A with ALMA at 1.2\,mm, corresponding to the continuum peak coordinates of A1 (green) and A2 (black). Except for CH$_3$OCH$_3$($13_{5,8}-13_{4,9}$), whose three blended lines are indicated with vertical blue lines, the Gaussian fits to all the lines observed in A2 are depicted in blue (see Table\,\ref{tlines}). Nearby molecular lines are labelled in red.}
\label{fspt}
\end{figure*}

\begin{figure}[!ht]
\centering
\includegraphics[scale=0.49]{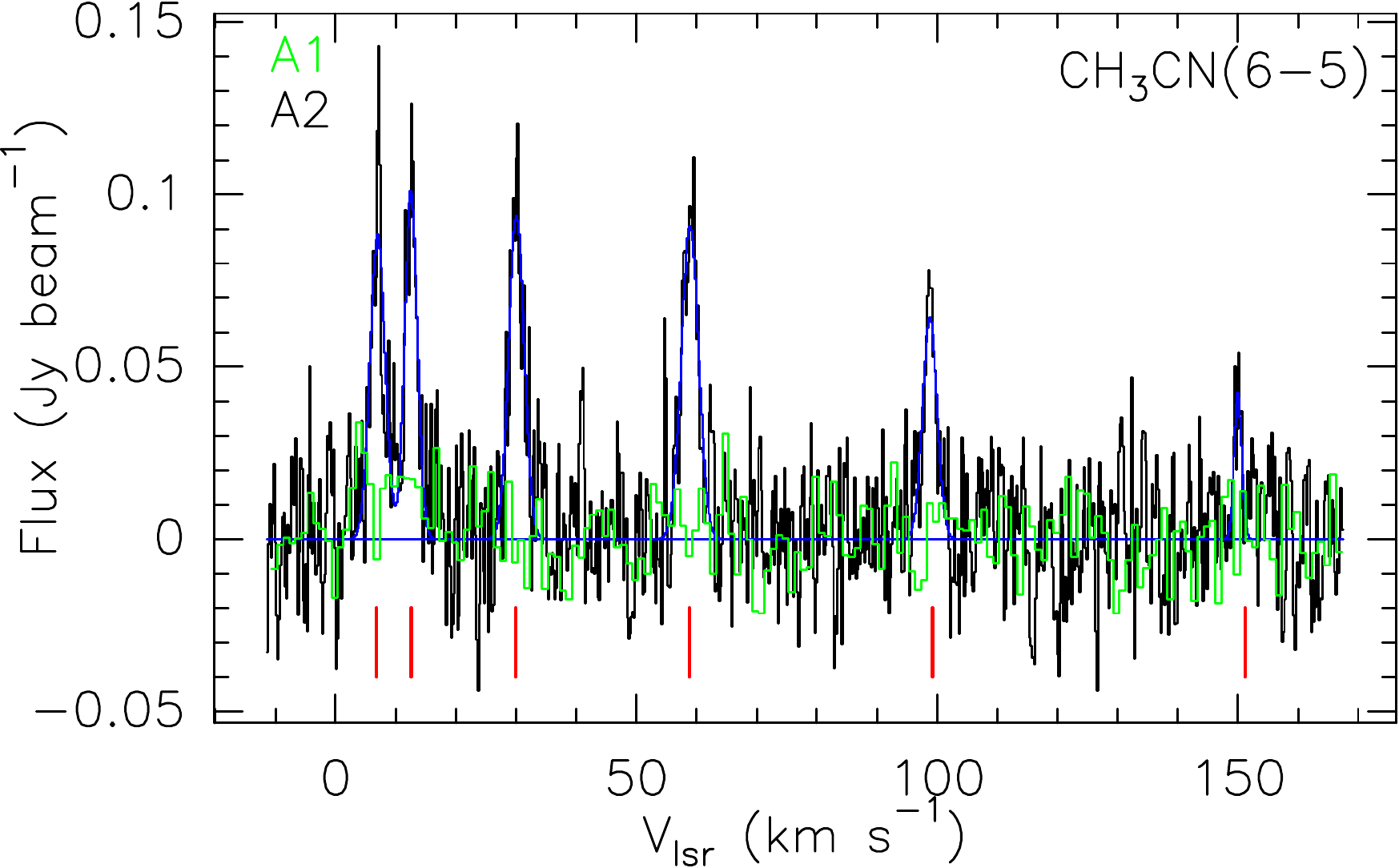}
\caption{CH$_3$CN($J = 6-5$), K = 0 -- 5 spectrum observed in IRAS\,4A with PdBI, corresponding to the continuum peak coordinates of A1 (green) and A2 (black). The spectrum of A1 was smoothed to a channel width of 1\,km\,s$^{-1}$. The Gaussian fits to the lines detected in A2 are depicted in blue (see Table\,\ref{tlines}). The positions of the different K transitions, from $K = 0$ (left) to $K = 5$ (right), are indicated in red.}
\label{fspt2}
\end{figure}

\begin{figure*}[!h]
\centering
\includegraphics[scale=0.7]{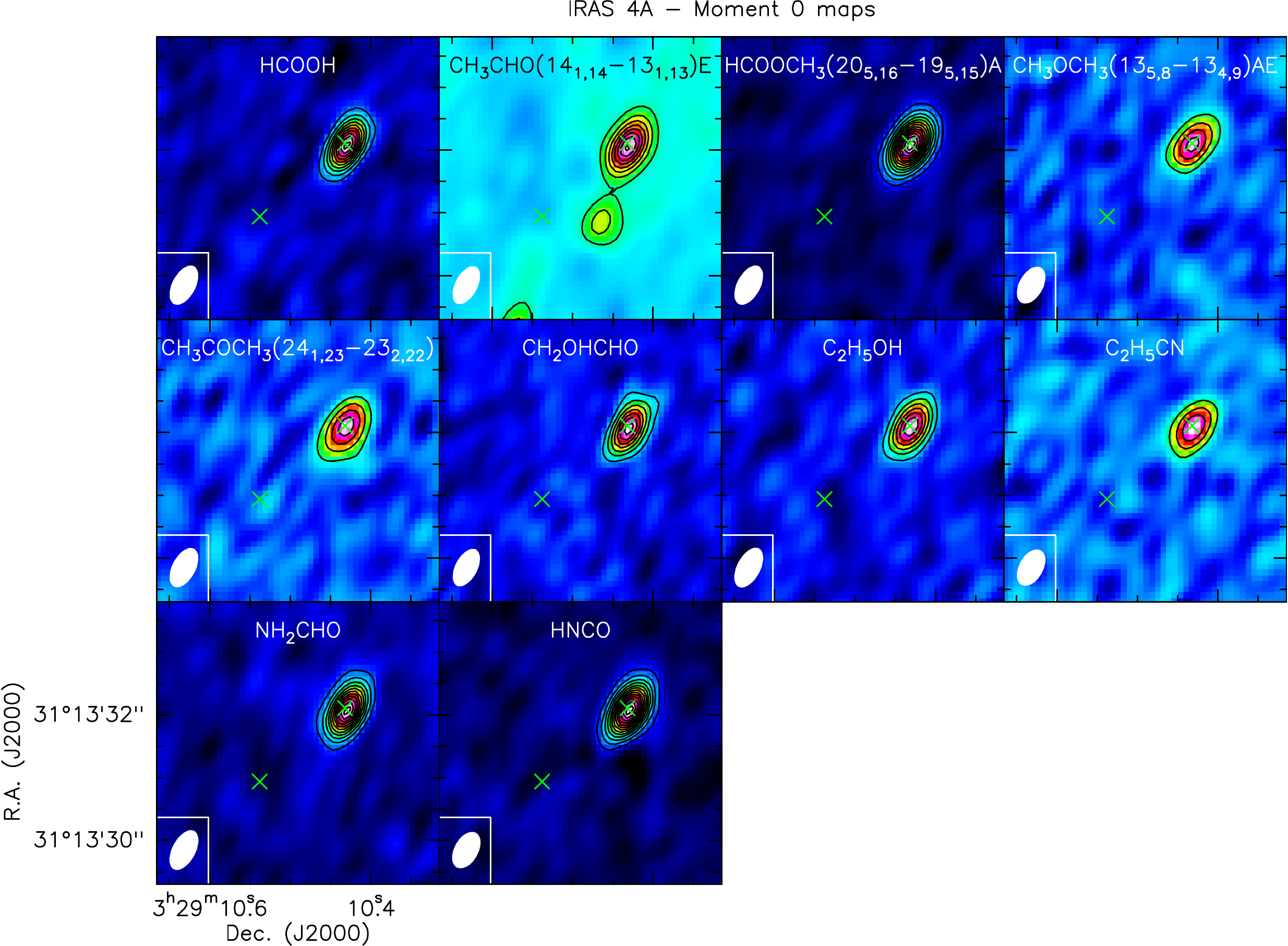}
\caption{Moment\,0 maps of representative molecular lines observed with ALMA. The transition is specified if more than one line from the same species was detected. The continuum peak position of the two protostellar cores A1 and A2 are indicated with green crosses. Contours start at 5$\sigma$ and increase by steps of 5$\sigma$ in all the maps. The values of $\sigma$ for each map are summarised in Table\,\ref{tmaps}. The synthesised beam is depicted in white in the lower left corner.}
\label{fmaps}
\end{figure*}

\begin{figure}[!h]
\centering
\includegraphics[scale=0.63]{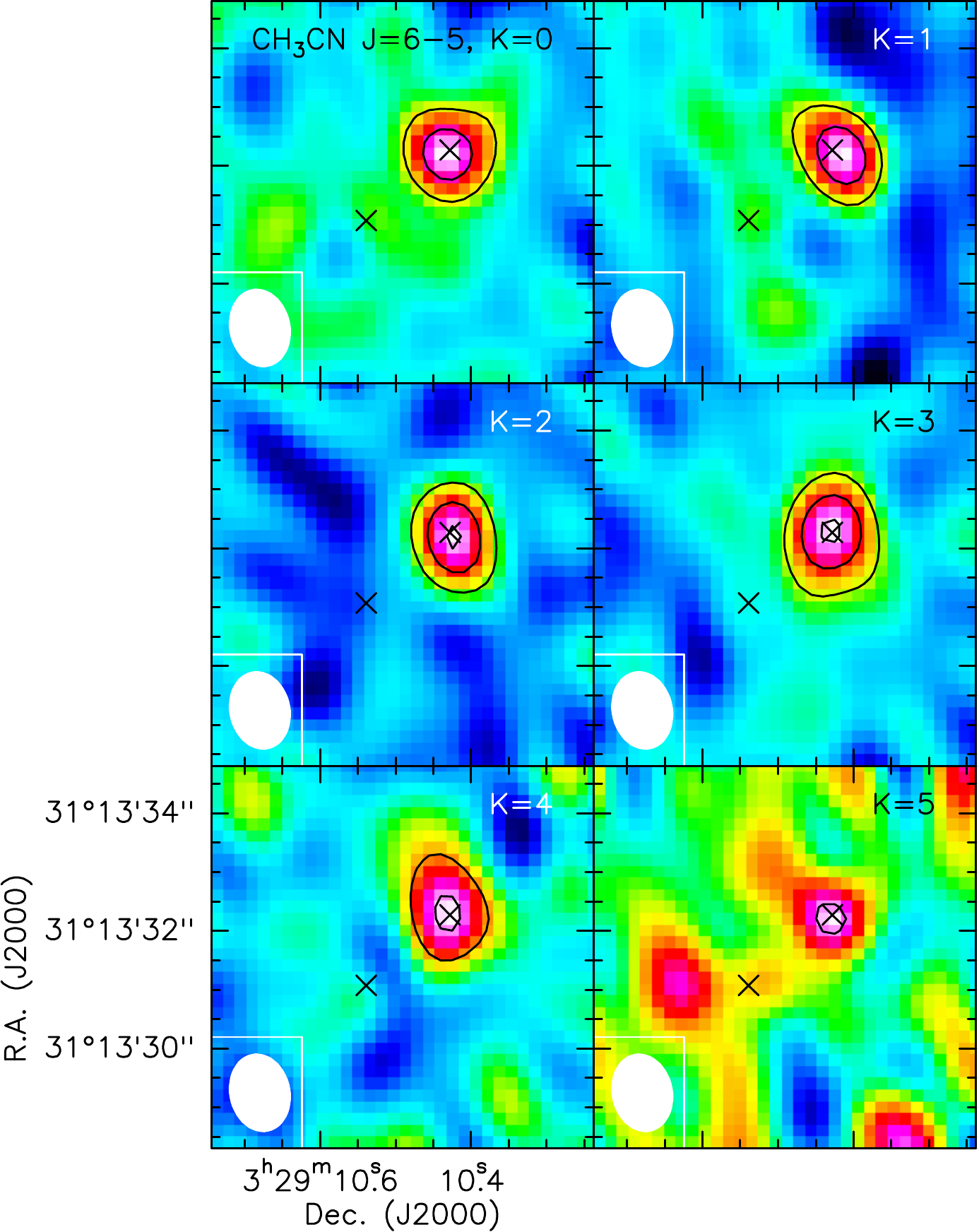}
\caption{Moment\,0 maps of the six CH$_3$CN ($J = 6-5$) lines observed with PdBI. The transition is specified in each panel. The continuum peak position of A1 and A2 are indciated with black crosses. Contours start at 4$\sigma$ and increase by steps of 4$\sigma$ in all the maps. The values of $\sigma$ for each map are summarised in Table\,\ref{tmaps}. The synthesised beam is depicted in white in the lower left corner.}
\label{fmaps2}
\end{figure}

Figures\,\ref{fspt} and \ref{fspt2} show the spectra of each line detected with ALMA and PdBI, respectively, at the continuum peak coordinates of A1 and A2. For each detected molecule with ALMA, moment\,0 maps of one representative line are shown in Fig.\,\ref{fmaps}. Moment 0 maps of the six CH$_3$CN lines detected with PdBI are presented in Fig.\,\ref{fmaps2}. The properties of these maps are summarised in Table\,\ref{tmaps}. From Figs.\,\ref{fspt} to \ref{fmaps2}, it is immediately evident that A2 emits strongly in all the identified organic species, whereas A1 appears devoid of them down to the sensitivity of our observations. While such a difference in molecular richness between the two objects was already known \citep[e.g][]{persson12,santangelo15}, this is the first time we are able to clearly disentangle the spatial emission from the two sources for such a large number of organic molecules. For CH$_3$CHO, the map also shows emission towards the south of both A1 and A2. This is due to contamination from the blue-shifted outflow lobe traced by SiO, whose systemic velocity is shifted by +14\,km\,s$^{-1}$ with respect to that of the CH$_3$CHO line.\footnote{The molecular outflows associated with IRAS\,4A will be the subject of a forthcoming paper.}. No contamination from SiO is expected on A2 itself, as the blue-shifted emission of this species is only detected well off the central protostellar core. The CH$_3$CN, $K = 5$ appears noisy because of the low S/N, barely above 4, of the peak emission around A2. The emission from all the identified organic molecules is very compact around A2, with deconvolved sizes of about 0.3$''$. Therefore, the hot corino surrounding A2 protostar has an approximate size of 70\,au. 

Through comparison with IRAM\,30 m data of IRAS\,4A obtained in the framework of the large programme ASAI (Astrochemical Surveys At IRAM\footnote{www.oan.es/asai}), we estimated the fraction of line flux recovered by the array for HNCO and HCOOCH$_3$ lines to be 88\% and 78\%, respectively. Thus, the bulk of the total emission for these two species is concentrated in the hot corino area. For CH$_3$CN, the fraction of flux recovered by the PdBI varies with $K$; it is around 30\% for $K = 0$ and 1 and 66\% for $K = 4$. Thus, a significant amount of flux in the lower energy CH$_3$CN lines, likely originating in the more extended protobinary envelope and/or the outflow(s), is filtered out by the array, while a considerably larger fraction of flux is recovered for the high energy transitions, indicating that the hot corino around A2 is the dominant contributor to the emission at higher $K$ levels. As for the other species, the lines identified in the present work remain undetected or very weakly detected down to the sensitivity of its observations. This further highlights the compact size of the emission, heavily diluted with the single dish.

\begin{table*}
\centering
\caption{Properties of the velocity-integrated molecular maps}
\begin{tabular}{llccc}
\hline
Molecule & Transition & RMS & Beam size & P.A.\\
  & & (mJy\,beam$^{-1}$\,km\,s$^{-1}$) & ($'' \times ''$) & ($^\circ$)\\
\hline
CH$_3$CN & $6_5-5_5$ & 13 & 1.35 $\times$ 1.04 & --168\\
CH$_3$CN & $6_4-5_4$ & 18 & 1.35 $\times$ 1.04 & --168\\
CH$_3$CN & $6_3-5_3$ & 27 & 1.35 $\times$ 1.04 & --168\\
CH$_3$CN & $6_2-5_2$ & 21 & 1.35 $\times$ 1.04 & --168\\
CH$_3$CN & $6_1-5_1$ & 21 & 1.35 $\times$ 1.04 & --168\\
CH$_3$CN & $6_0-5_0$ & 26 & 1.35 $\times$ 1.04 & --168\\
\hline
t-HCOOH & $12_{0,12}-11_{0,11}$ & 7 & $0.68 \times 0.37$ & --28\\ 
CH$_3$CHO & $14_{1,14}-13_{1,13}$\,E & 16 & $0.66 \times 0.35$ & --28\\ 
HCOOCH$_3$ & $20_{5,16}-19_{5,15}$\,A & 6 & $0.68 \times 0.37$ & --28\\ 
CH$_3$OCH$_3$ & $13_{5,8}-13_{4,9}$\,AE & 5 & $0.64 \times 0.38$ & --29\\ 
(CH$_3$)$_2$CO & $24_{1,23}-23_{2,22}$\,AA & 5 & $0.68 \times 0.37$ & --28\\ 
cis-CH$_2$OHCHO & $25_{2,24}-24_{1,23}$ & 6 & $0.65 \times 0.35$ & --28\\ 
CH$_3$CH$_2$OH & 14$_{3,11}-13_{3,10}$ & 6 & $0.69 \times 0.37$ & --28\\
CH$_3$CH$_2$CN & $29_{2,27}-28_{2,26}$ & 5 & $0.63 \times 0.38$ & --29\\ 
NH$_2$CHO & $12_{0,12}-11_{0,11}$ & 7 & $0.69 \times 0.37$ & --28\\ 
HNCO & $12_{0,12}-11_{0,11}$ & 12 & $0.63 \times 0.38$ & --29\\ 
\hline
\end{tabular}
\label{tmaps}
\end{table*}

\subsection{Derivation of column densities and fractional abundances}\label{deriv}

\subsubsection{IRAS\,4A2}\label{a2}

\begin{figure}[!h]
\centering
\includegraphics[scale=0.51]{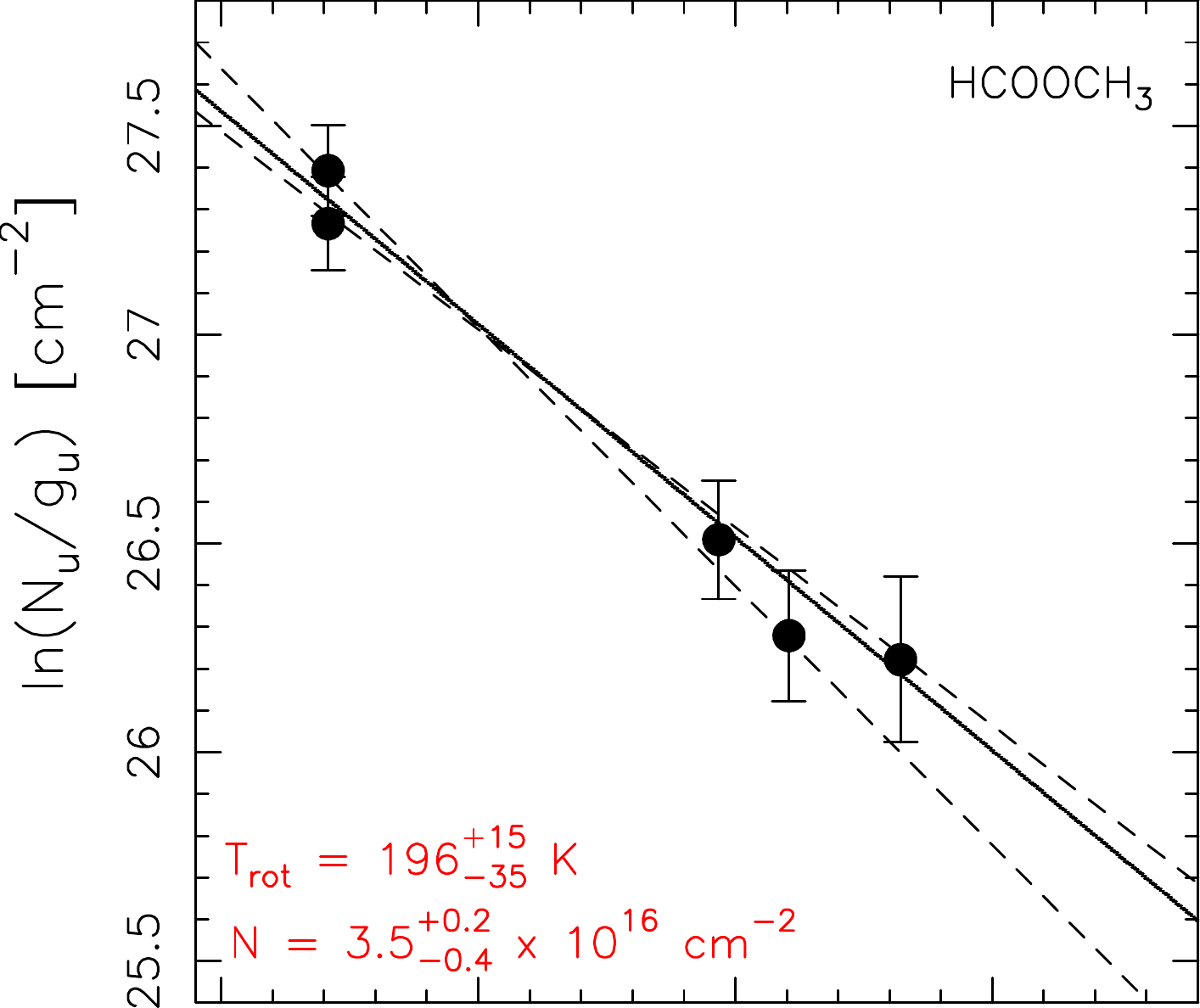}

\includegraphics[scale=0.51]{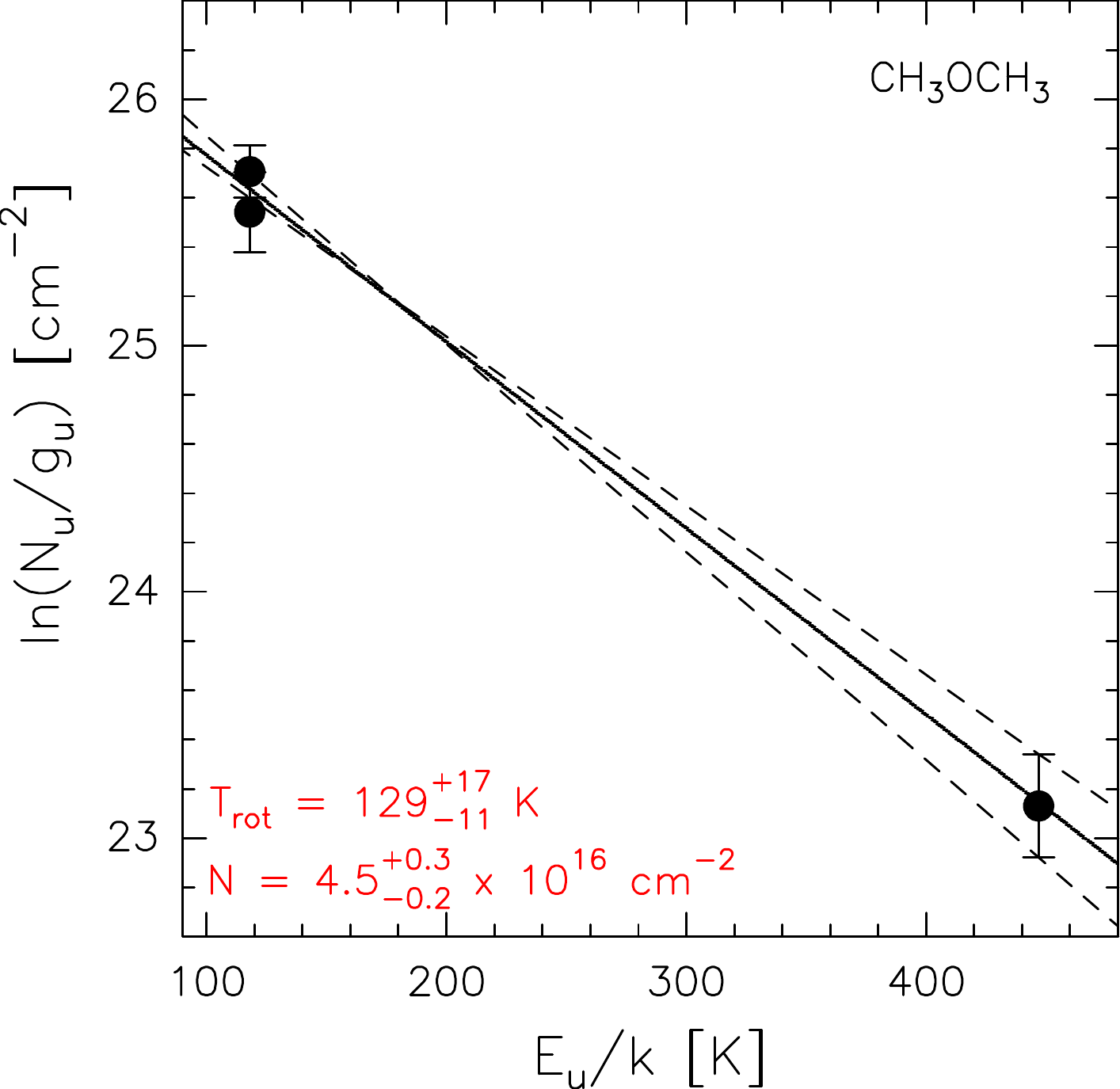}
\caption{Rotational diagrams of methyl formate (HCOOCH$_3$) and dimethyl ether (CH$_3$OCH$_3$) for IRAS\,4A2. Data points are depicted by black circles. Error bars account for the spectral RMS and the calibration error. The solid straight lines correspond to the best fit to the data points. Each pair of dashed straight lines represent the opposing limits encompassing the whole range of possible solutions within the error bars. The best fit rotational temperatures and molecular column densities are indicated in red.}
\label{frd}
\end{figure}

In A2, we detect more than two lines from HCOOCH$_3$ and CH$_3$OCH$_3$. In order to estimate the excitation conditions of these two molecular species, we used the rotational diagram approach, which assumes local thermodynamic equilibrium (LTE) and optically thin line emission. We verified the latter condition to be true a posteriori: the optical depth values we measured are of the order of 0.1 or well below. Under these assumptions, a single temperature, known as \textit{rotational temperature}, $T_\mathrm{rot}$, describes the relative population distribution of each energy level of a given molecule, as follows:

\begin{equation}
N_\mathrm{u} = \frac{3}{8} \frac{k g_\mathrm{u}}{\pi^3 \nu S \mu^2} \frac{1}{\eta_\mathrm{bf}} \int T_\mathrm{B} dV,
\label{enu}
\end{equation}

\begin{equation}
\ln \frac{N_\mathrm{u}}{g_\mathrm{u}} = \ln N-\ln Q(T_\mathrm{rot})-\frac{E_\mathrm{u}}{k T_\mathrm{rot}},
\label{erd}
\end{equation}
where $N_\mathrm{u}$ is the column density of the upper energy level of the transition, $k$ Boltzmann's constant, $g_\mathrm{u}$ the upper level degeneracy, $\nu$ the frequency of the transition, $S$ the line strength, $\mu$ the dipole moment, $\int T_\mathrm{B} dV$ the velocity-integrated line flux, $N$ the total molecular column density, $Q(T_\mathrm{rot})$ the partition function, and $E_\mathrm{u}$ the energy of the upper energy level. The partition function takes into account the vibrationally excited states for HCOOCH$_3$. Finally, the beam filling factor is given by

\begin{equation}
\eta_\mathrm{bf} = \frac{\theta_\mathrm{hc}^2}{\theta_\mathrm{hc}^2+\theta_\mathrm{beam}^2},
\label{ebf}
\end{equation}
with the deconvolved hot corino size, $\theta_\mathrm{hc}$, roughly equal to 0.3$''$, and $\theta_\mathrm{beam}$ the beam size of the observations. The resulting rotational diagrams are presented in Fig.\,\ref{frd}, where the error bars on each data point include the RMS and calibration uncertainties. In order to account for these, each rotational diagram shows, in addition to the best fit line, two opposite lines that encompass the whole range of possible rotational temperatures and molecular column densities within the errors of all the data points. By doing so, we obtain $T_\mathrm{rot} = 196^{+15}_{-35}$\,K and $N = 3.5^{+0.2}_{-0.4} \times 10^{16}$\,cm$^{-2}$ for HCOOCH$_3$, and $T_\mathrm{rot} = 132^{+17}_{-11}$\,K and $N = 4.5^{+0.3}_{-0.2} \times 10^{16}$\,cm$^{-2}$ for CH$_3$OCH$_3$ (see Table\,\ref{tcol}).

Based on the derived rotational temperatures for these two species, which are consistent with typical hot corino conditions, we adopted the excitation temperature range $T_\mathrm{ex} = 100 - 200$\,K to estimate the column densities of the other eight molecular species, for which we detect either one or two relatively isolated transitions. Once more, we assumed LTE and optically thin line emission, and we took the RMS and calibration errors in the line fluxes into account. The resulting column densities are listed in Table\,\ref{tcol} and they span values of the order of $10^{15}$ to $10^{16}$\,cm$^{-2}$.

As stated above, the fluxes of the lower $K$ CH$_3$CN($J = 6-5$) lines are largely filtered out by the PdBI, and they also display signs of self-absorption, which altogether indicates the presence of CH$_3$CN in the outer envelope. An attempt to perform a rotational diagram fit thus yielded unrealistically high rotational temperatures. On the other hand, the $K = 5$ CH$_3$CN line has a low S/N, and is possibly blended with a CH$_2$OHCHO line. We therefore used the $K = 4$ line to estimate the column density of this species within the range $T_\mathrm{ex} = 100 - 200$\,K.

\subsubsection{IRAS\,4A1: Upper limits to molecular column densities}\label{a1}

Unlike A2, A1 displays no detectable line emission from any of the organic molecules considered in this work. In order to study the differences in molecular richness present between these two protostellar cores, we provide, in Table\,\ref{tcol}, 3$\sigma$ upper limits to the beam-averaged column densities in A1, where $\sigma$ is the RMS measured after smoothing the spectra to a channel width of 1\,km\,s$^{-1}$. We assume for simplicity that the dust emission at the frequencies of our observations is optically thin and therefore that the lack of line emission from COMs in A1 is not due to its complete attenuation by foreground dust. As discussed in Sect.\,\ref{cont}, the effect of dust optical depth in A2 is likely weak or negligible. A more detailed discussion about dust optical depth and how it may affect our results, particularly in A1, can be found in Sect.\,\ref{da1}. Based on the typical values measured for A2, we adopted a line width of 2.5\,km\,s$^{-1}$ for all the species. We considered two different excitation temperatures: $T_\mathrm{ex} = 100$\,K for comparison with A2, and $T_\mathrm{ex} = 50$\,K, as the apparent lack of organic molecular emission lines in the gas around the protostar A1 suggests a temperature lower than 100\,K along most of the observed dust column. We obtained the upper limit column densities assuming LTE conditions.

CH$_3$CN(6$_0-5_0$) is tentatively detected in the direction of A1 at $S/N \sim 4$. The fact that only the $K = 0$ transition is weakly detected while the higher $K$ transitions are not detected suggests that the emission from this molecule originates in the outer envelope, rather than in a hypothetical hot corino. The upper limits to the CH$_3$CN column density shown in Table\,\ref{tcol} correspond to the $K = 4$ transition.

\begin{table*}
\centering
\caption{Molecular column densities and abundances with respect to H$_2$}
\begin{tabular}{lcccccccc}
\hline
 & \multicolumn{3}{c}{IRAS\,4A2} & & \multicolumn{4}{c}{IRAS\,4A1$^\mathrm{a}$}\\
 \cline{2 - 4} \cline{6 - 8}
Molecule & $T_\mathrm{rot}$$^\mathrm{b}$ & $N$$^\mathrm{b}$ & $X_\mathrm{A2}$ & & $N$(50\,K) & $N$(100\,K) & $X_\mathrm{A1}$ & $X_\mathrm{A2}/X_\mathrm{A1}$\\
 & (K) & (cm$^{-2}$) & & & (cm$^{-2}$) & (cm$^{-2}$) & &\\
\hline
t-HCOOH & 100 -- 200 & $(4.6 - 8.3) \times 10^{15}$ & (0.6 - 2.9) $\times 10^{-9}$ & & $< 1.0 \times 10^{14}$ & $< 1.3 \times 10^{14}$ & $<$ 9.7 $\times 10^{-12}$ & > 62\\
CH$_3$CHO & 100 -- 200 & $(1.0 - 1.9) \times 10^{16}$ & (1.1 - 7.4) $\times 10^{-9}$ & & $< 1.5 \times 10^{14}$ & $< 1.7 \times 10^{14}$ & $<$ 1.3 $\times 10^{-11}$ & > 85\\
HCOOCH$_3$ & 196$^{+15}_{-35}$ & 3.5$^{+0.2}_{-0.4} \times 10^{16}$ & 1.1$^{+0.2}_{-0.1} \times 10^{-8}$ & & $< 1.4 \times 10^{15}$ & $< 1.1 \times 10^{15}$ & $<$ 8.2 $\times 10^{-11}$ & > 120\\
CH$_3$OCH$_3$ & 129$^{+17}_{-11}$ & 4.5$^{+0.3}_{-0.2} \times 10^{16}$ & 1.0$^{+0.1}_{-0.1} \times 10^{-8}$ & & $< 6.5 \times 10^{15}$ & $< 5.6 \times 10^{15}$ & $<$ 4.3 $\times 10^{-10}$ & > 21\\
(CH$_3$)$_2$CO$^\mathrm{c}$ & 100 -- 200 & $(4.2 - 5.9) \times 10^{15}$ & (0.5 - 2.2) $\times 10^{-9}$ & & $< 6.4 \times 10^{14}$ & $< 3.8 \times 10^{14}$ & $<$ 2.9 $\times 10^{-11}$ & > 17\\
cis-CH$_2$OHCHO & 100 -- 200 & $(5.8 - 6.8) \times 10^{15}$ & (0.7 - 2.4) $\times 10^{-9}$ & & $< 5.4 \times 10^{14}$ & $< 2.8 \times 10^{14}$ & $<$ 2.1 $\times 10^{-11}$ & > 33\\
C$_2$H$_5$OH & 100 -- 200 & (3.3 - 4.9) $\times 10^{16}$ & (0.4 - 1.9) $\times 10^{-8}$ & & $< 2.1 \times 10^{15}$ & $< 1.6 \times 10^{15}$ & $<$ 1.2 $\times 10^{-10}$ & > 33\\
C$_2$H$_5$CN & 100 -- 200 & $(1.4 - 1.5) \times 10^{15}$ & (1.7 - 5.6) $\times 10^{-10}$ & & $< 2.8 \times 10^{14}$ & $< 1.1 \times 10^{14}$ & $<$ 1.0 $\times 10^{-11}$ & > 17\\
NH$_2$CHO & 100 -- 200 & $(1.0 - 2.0) \times 10^{15}$ & (1.2 - 6.7) $\times 10^{-10}$ & & $< 2.1 \times 10^{13}$ & $< 2.8 \times 10^{13}$ & $<$ 2.1 $\times 10^{-12}$ & > 57\\
HNCO & 100 -- 200 & $(2.3 - 4.0) \times 10^{15}$ & (0.3 - 1.4) $\times 10^{-9}$ & & $< 4.5 \times 10^{13}$ & $< 5.5 \times 10^{13}$ & $<$ 4.2 $\times 10^{-12}$ & > 71\\
CH$_3$CN$^\mathrm{d}$ & 100 -- 200 & (1.9 - 4.8) $\times 10^{16}$ & (0.4 - 2.1) $\times 10^{-8}$ & & $<$\,6.4 $\times 10^{14}$ & $<$\,4.8 $\times 10^{14}$ & $<$\,2.7 $\times 10^{-11}$ & > 150\\
\hline
\end{tabular}
\\
\raggedright $^\mathrm{a}$ 3$\sigma$ upper limits, where $\sigma$ is the RMS measured after smoothing the spectra to a channel width of 1\,km\,s$^{-1}$.\\
$^\mathrm{b}$ For HCOOCH$_3$ and CH$_3$OCH$_3$, $T_\mathrm{rot}$ and $N$ correspond to the best fit results from the rotational diagram analysis. For the other molecules, they correspond to the indicated range of excitation temperatures (see text).\\
$^\mathrm{c}$ Derived from $24_{1,23}-23_{2,22}$\,AA transition.\\
$^\mathrm{d}$ Derived from the $K = 4$ transition.
\label{tcol}
\end{table*}

\subsubsection{Molecular abundances}\label{abu}

CH$_3$OH is often used as a reference to describe relative molecular abundances. As this molecule was not observed in our spectral set-up, we here adopt dust emission as a reference. Table\,\ref{tcol} summarises the resulting molecular abundances, $X = N/N_\mathrm{H_2}$, associated with the two cores, obtained from the values listed in Tables\,\ref{tmass} and \ref{tcol}. The ranges of values given for A2 correspond to a range of excitation and dust temperatures between 100 and 200\,K, where we assume that $T_\mathrm{ex} = T_\mathrm{d}$. For A1, we adopted the temperature that yielded the most conservative upper limit. This is 100\,K for all the molecular species but C$_2$H$_5$CN, for which it is 50\,K. The last column of Table\,\ref{tcol} lists the abundance ratios between A2 and A1.

It is important to notice that, while $N_\mathrm{H_2}$ is a beam-averaged quantity, the molecular column densities derived for A2 are corrected for beam dilution, as the hot corino is slightly smaller than the beam. The abundances for A2 are therefore underestimated, as only a fraction of the peak dust continuum emission from which we computed $N_\mathrm{H_2}$ is likely associated with the hot corino area itself.

\section{Discussion}\label{discussion}

\subsection{IRAS\,4A2: Typical hot corino abundances}\label{da2}

While the molecular column densities listed in Table\,\ref{tcol} for A2 are comparable to those derived by \citet{taquet15} for the seven molecular species common to the two studies, our molecular abundances tend to be higher than theirs by about one order of magnitude. This is most likely due to the coarser angular resolution of the observations analysed by \citet{taquet15} with respect to ours. Indeed, their 2$''$ beam is insufficient to disentangle properly the two cores, whose angular separation is 1.8$''$. Since COMs are detected only in A2, but most of the continuum emission comes from A1, the averaged H$_2$ column density they estimate includes a non-negligible amount of gas from A1 that is cooler and devoid of COMs. This results in lower overall abundances for IRAS\,4A as a whole.

To our knowledge, there are only three other hot corino sources for which interferometric abundance measurements of multiple COMs exist in the literature. These are the protobinary I16293 in Ophiucus, the newly discovered hot corino in the Bok globule B335, and IRAS\,2A in Perseus. The top panel of Fig.\,\ref{fabu} plots the molecular abundances we derived for IRAS\,4A2 together with those of the three mentioned hot corinos. We stress that only interferometric measurements are shown for consistency with our work. This plot shows that A2 contains typical hot corino abundances that are comparable to those measured in the other three objects.

\subsection{IRAS\,4A1: Possible explanations for the lack of COM emission}\label{da1}

In Sects.\,\ref{a1} and \ref{abu}, we computed beam-averaged upper limits to the molecular column densities and abundances in A1, the close companion of the hot corino source A2, which despite emitting strongly in dust continuum (see Fig.\,\ref{fcont}) and driving a powerful molecular outflow \citep{santangelo15}, displays no detectable emission in COMs or even in simpler organic species such as HNCO.

\begin{figure}[!h]
\centering
\begin{tabular}{c}
\includegraphics[scale=0.58]{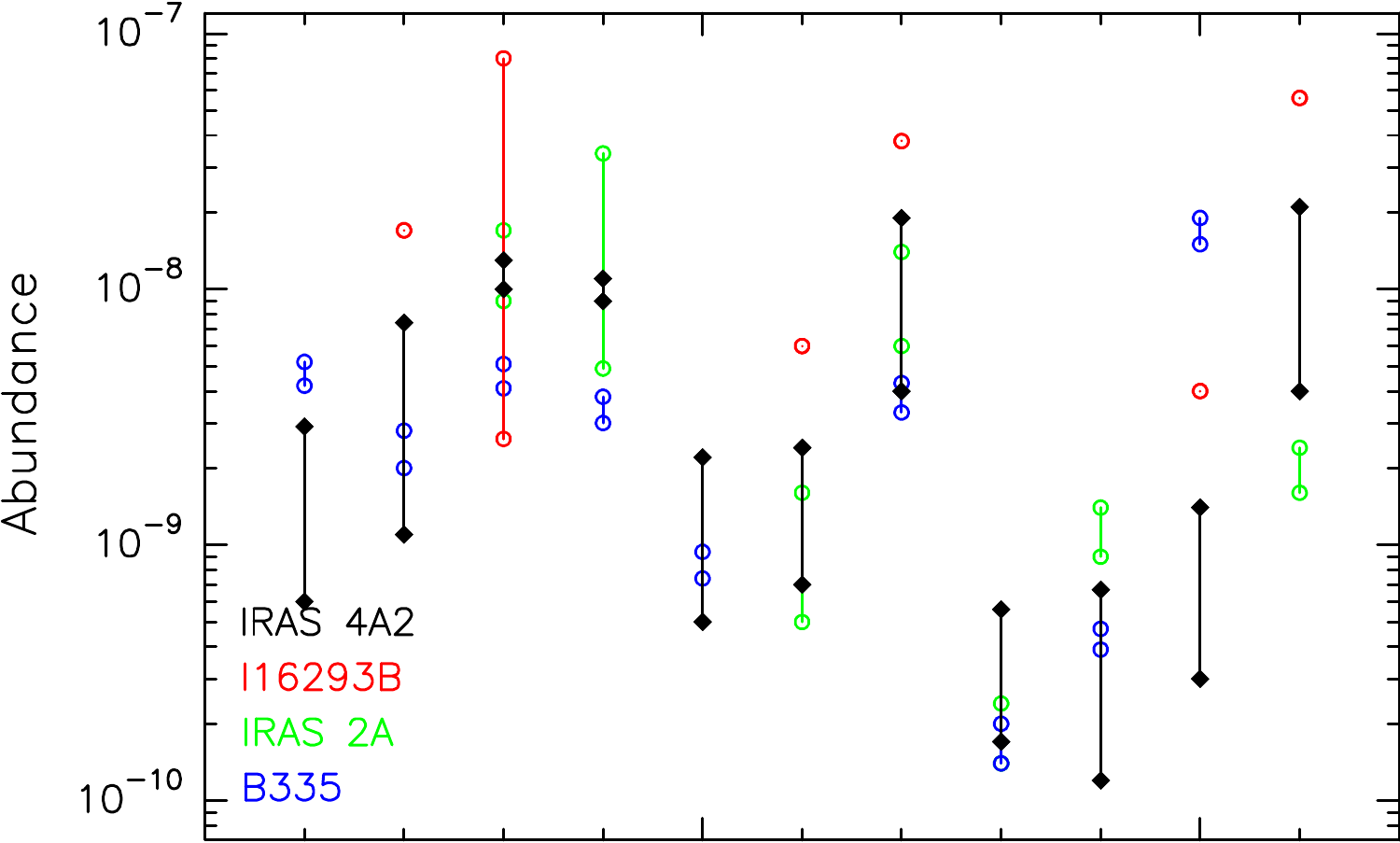}\\
\includegraphics[scale=0.58]{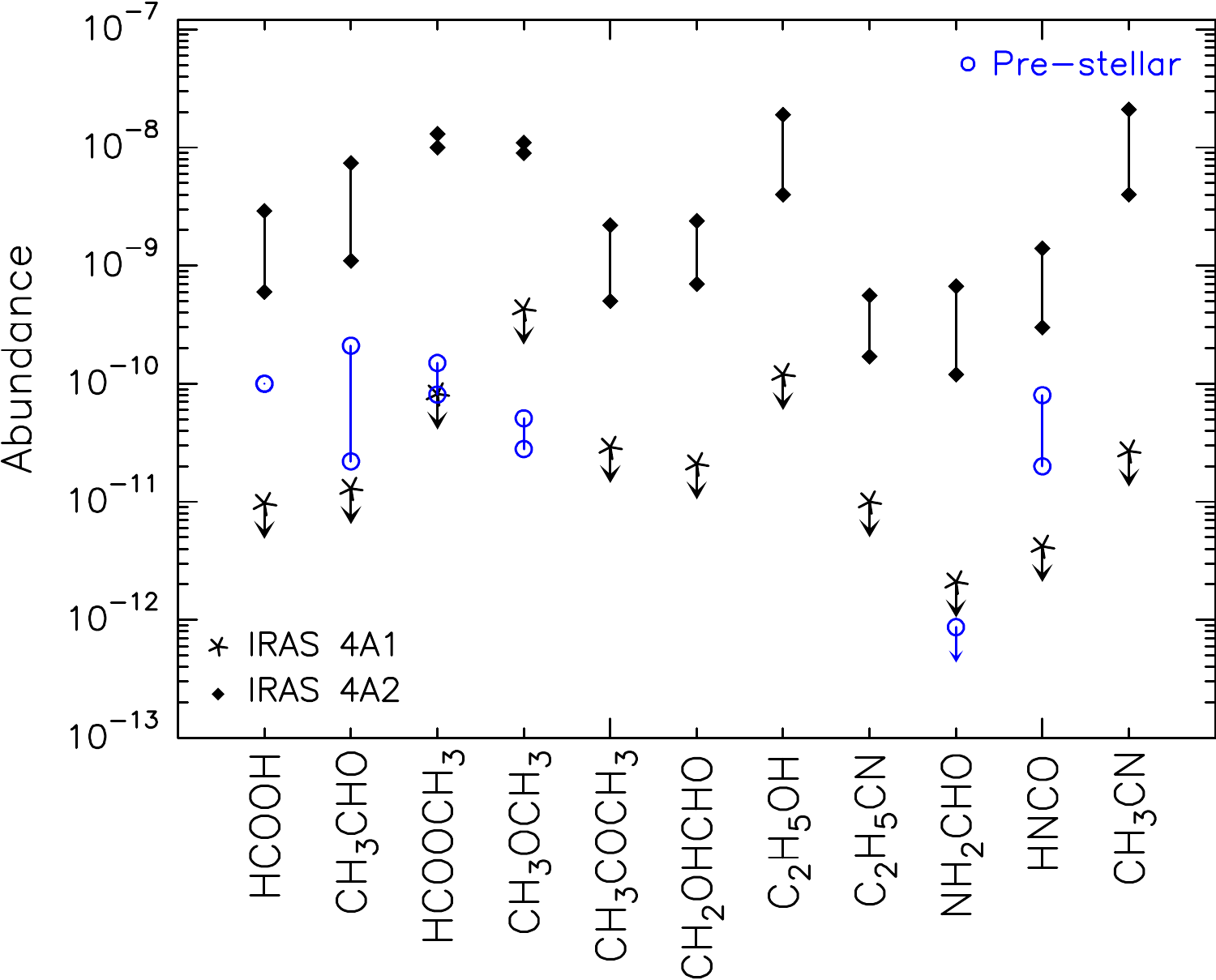}
\end{tabular}
\caption{\textit{Top}: Molecular abundances with respect to H$_2$ for the following hot corinos: IRAS\,4A2 (black; this work), I16293B (red), IRAS\,2A (green), and B335 (blue). All the values are derived from interferometric observations. References: I16293B: \citet{bisschop08,jorgensen12}; IRAS\,2A: \citet{persson12,taquet15}; B335: \citet{imai16}. \textit{Bottom}: Molecular abundances with respect to H$_2$ derived in this work for IRAS\,4A2 and IRAS\,4A1 (upper limits). Blue symbols depict the values recently reported for the prestellar core L1544 by \citet{vastel14,yo15}, and \citet{js16}.}
\label{fabu}
\end{figure}

The bottom panel of Fig.\,\ref{fabu} shows a comparison between the abundances we derive for A1 and A2. As can be seen in this plot, and also in Table\,\ref{tcol}, the difference in COMs abundances between the two objects is larger than a factor 17 for all the molecular species, rising up to a value higher than 100 for HCOOCH$_3$ and CH$_3$CN. This huge difference in chemical richness between two neighbouring protostellar cores that are formed from the same parental molecular clump is puzzling and deserves further investigation. We here discuss three different possibilities that may explain our results.

The first possibility is that A1 contains a hot corino, but it is heavily attenuated by dust. Indeed, the dust emission we detect in A1 is likely optically thick, at least at 1.2\,mm. Let us assume that  A1 actually hosts a hot corino similar in size ($\sim$\,70\,au) and molecular richness to that in A2, but that the dust around A1 protostar is sufficiently optically thick to block the emission completely from the molecular species considered in this study. If this is the case, we can roughly estimate the dust optical depth necessary to attenuate line emission from COMs to the level of non-detection. At 2.7\,mm, where such effects should be minor, if not negligible, we consider the CH$_3$CN($J = 6-5$) $K = 4$ line, which is less affected by effects of interferometric filtered emission than lower $K$ lines. If this line were to be as intense in A1 as that in A2, it should suffer an attenuation of $\sim$80\% by dust in the former in order to fall below 3$\sigma$ (for a channel width of 1\,km\,s$^{-1}$). This translates into a dust optical depth $\sim 1.6$ (a similar exercise at 1.2\,mm taking CH$_3$CHO yields $\tau \sim 2.3$). This optical depth would correspond to the \emph{foreground} dust only to attenuate completely the molecular line emission that lies behind, which implies a total optical depth of at least twice as much for the whole core, i.e. $\tau > 3.2$. In other words, the total H$_2$ column density for A1 should be higher than $2.7 \times 10^{26}$\,cm$^{-2}$ within the 1$''$ beam of the 2.7\,mm observations, which translates into a beam-averaged density of $\sim 10^{11}$\,cm$^{-3}$. Similar results are obtained if we consider CH$_3$CHO at 1.2\,mm instead. While such a large gas density is not totally impossible, it is highly unlikely. Moreover, Choi et al. (2007) reported evidence that NH$_3$ lines are weaker in A1 than in A2 at 1.3\,cm, with a NH$_3$-to-dust ratio seven times larger in the latter than in the former. At these wavelengths, dust is most likely optically thin. In summary, it is reasonable to assume that the difference in molecular line emission between A1 and A2 reflects a real difference in chemical richness and/or hot corino sizes and that dust optical depth does not play a major role. In what follows, we assume that this is the case.

This leaves us with the possibility that that A1 hosts no hot corino whatsoever, or alternatively, that it contains a hot corino whose size is so small that its emission is beam diluted, even at the angular resolution of our observations ($\sim 0.5''$). If the latter is the case, we can roughly estimate how small this hot corino should be to contain COMs in abundances comparable to those in IRAS\,4A2, by considering a comparable excitation temperature and column density of, for example CH$_3$CN($J = 6-5$) $K = 4$, which is less unlikely to suffer from dust optical depth effects than any of the lines observed at 1.2\,mm. The assumption of a similar column density is rather conservative, as one would expect a much lower value for a much smaller hot corino. With these assumptions, a hypothetical hot corino in A1 should have a size smaller than about 12\,au. 

\subsection{IRAS\,4A1: Abundances below prestellar values?}\label{prestellar}

For comparison with A1, Fig.\,\ref{fabu} (\textit{bottom}) plots molecular abundances measured in a prestellar core, L1544, as reported in \citet{vastel14,yo15}, and \citet{js16}. The discovery of COMs in prestellar cores challenged the most accepted scenario of grain surface astrochemical models \citep[e.g.][]{garrod08}. According to this scenario, COMs form on the surface of icy mantles coating dust grains via recombination of radicals during the warm-up phase of protostellar collapse ($T > 30$\,K). Once the sublimation temperature of water ice is attained ($T > 100$\,K), the COMs thus formed are released into the gas, where they become detectable. Given that prestellar cores have not yet experienced such a warm-up phase, the presence of COMs in this type of objects must have a different origin. Non-thermal processes have been proposed, including photo-desorption by cosmic rays and secondary UV photons \citep{hasegawa93} and reactive desorption \citep{vasyunin13}. More recently, efficient cold gas phase reactions have also been invoked to explain the presence of COMs in cold gas \citep[e.g.][]{balucani15}.

A brief inspection of the bottom panel in Fig.\,\ref{fabu} strikingly reveals that the upper limits to the abundances measured in A1, with the exception of dimethyl ether (CH$_3$OCH$_3$), are lower than those observed in L1544 by up to about one order of magnitude. In other words, the protostellar core A1 appears to be more deficient in COMs than the prestellar core L1544. This still holds if we consider instead the cold core L1689B, for which \citet{bacmann12} tentatively derived CH$_3$CHO, HCOOCH$_3$, and CH$_3$OCH$_3$ abundances of a few times $10^{-10}$.

It is likely that COMs are present in the cold outer regions of gas surrounding the protobinary with abundances comparable to those measured in L1544 \citep[see e.g.][]{jaber14}, which is the reason why we plot the abundances of L1544 in the bottom panel of Fig.\,\ref{fabu}. However, the emission in the outer envelope may be too extended and thus resolved out by the interferometer, or it may simply be too weak to be detectable at the cold gas temperatures involved. Some of the molecules studied in this work, including CH$_3$CHO, HCOOCH$_3$ and (CH$_3$)$_2$CO, display weak absorption lines in the direction of A1 (see Fig.\,\ref{fspt}), and the $K = 0$ line of the CH$_3$CN(J = 6-5) transition is marginally detected and possibly self-absorbed, which altogether indicates that COMs are indeed present in the more external envelope. \citet{vastel14} found that the region containing detectable amounts of COMs in L1544 is located at the border of the core, at a radius of $\sim 8000$\,AU, i.e. an area that is less dense and more exposed to external ultraviolet and cosmic ray irradiation than the inner, denser regions of the core. More recently, \citet{js16} confirmed that lower density gas regions in prestellar cores are richer in COMs than higher density gas regions. One may argue that this is a natural consequence of the balance between adsorption and desorption of COMs to and from dust grains: under equilibrium, the number density of COMs are constant regardless of the H$_2$ density \citep[e.g.][]{aikawa05,soma15}, and hence, their fractional abundances likely decrease in the densest regions. This could in principle explain the lower abundances in A1 with respect to those measured in the outer envelopes of pre- and protostellar cores, as the H$_2$ density is significantly higher in the A1 core ($\sim$$10^{8}$\,cm$^{-3}$; see Sect.\,\ref{cont}). However, this argument should be considered with caution, as it assumes that COMs are synthesised exclusively on the surface of dust grains and that they do not undergo reactions in the gas phase. This is unlikely, at least for some COMs \citep{balucani15,barone15,er16}.

In this context, A1 may represent an intermediate phase between a prestellar core and a hot corino, where infall and ejection of material in the form of bipolar outflows are already in place, but the gas immediately surrounding the central protostar is sufficiently cold and dense for either COMs or their precursors to be heavily frozen onto the mantles of dust grains in an environment that is well shielded from external ultraviolet/cosmic ray radiation.

Whether there is an undetected hot corino or no hot corino whatsoever in A1, our results point towards an enormous difference in molecular richness between A1 and A2. This may be due to protostar A2 being either more massive or subject to a higher accretion rate (or both) than A1. In other words, the prestellar core from which the IRAS\,4A system originated might have fragmented in an inhomogeneous manner, which would result in the core leading to A2 being initially more massive and thus undergoing higher accretion rate. This would explain the smaller current envelope mass in A2 with respect to A1 and the dynamically older molecular outflow in A2 with respect to that driven by A1 \citep{santangelo15}.
%

\subsection{Comparison with the protostellar binary IRAS\,16293--2422}

It is worth comparing IRAS\,4A with I16293, which is another well-studied protostellar binary hosting hot corino activity. I16293 has been the subject of numerous observations aiming to study the presence of COMs and their spatial distribution \citep[e.g.][]{cazaux03,bottinelli04a,caux11,oya16}. This object is composed of sources A and B, separated by 5.1$''$, or 620\,au at a distance of 120\,pc \citep{looney00,chandler05}. As in the case of IRAS\,4A, the weakest millimetre continuum source (i.e. source A) is the richest in COMs \citep[e.g.][]{jorgensen11,jorgensen16}.

There is, however, a substantial difference between I16293 and IRAS\,4A: the latter does not display \textit{any} detectable signs of COMs or simpler related molecules in A1, while the former does emit lines from some COMs and other molecules in source B. Indeed, while I16293A contains a larger variety of COMs than I16293B \citep[e.g.][]{jorgensen11}, including C$_2$H$_5$OH, HCOOH, and notably N-bearing COMs, source B competes with source A when it comes to the contents of a number of O-bearing COMs, such as CH$_3$CHO, CH$_3$OCH$_3$, HCOOCH$_3$, and (CH$_3$)$_2$CO. As shown in Fig.\,\ref{fabu}, the abundances of these molecules in source B are comparable to those of other known hot corinos, including IRAS\,4A2. Moreover, the differences in the abundances of COMs between sources A and B in I16293 do not exceed one order of magnitude \citep[e.g.][]{bisschop08}, while they are significantly larger between A1 and A2 in IRAS\,4A.

In conclusion, while I16293 and IRAS\,4A are similar in that they both show striking differences in molecular richness between the two protostellar cores they host, I16293 already contains hot corino activity in both A and B, while IRAS\,4A harbours only one detectable hot corino, i.e. A2. One might speculate that IRAS\,4A is a younger version of I16293 and that it may take time to properly develop a hot corino around a protostar. Indeed, the modelling results reported by \citet{aikawa12} indicate that chemical complexity increases during the evolution from a dense molecular core to a protostellar object. Alternatively, both IRAS\,4A and I16293 may be in a similar evolutionary phase, but IRAS\,4A1 might have been in a low-accretion stage for a sufficiently long time to suppress or considerably weaken any hot corino activity that may have previously existed in this source.

\section{Conclusions}\label{conclusions}

We carried out an interferometric study of multiple organic molecules in the protostellar binary NGC1333\,IRAS\,4A, using ALMA at 1.2\,mm and an angular resolution of $\sim$\,0.5$''$, and IRAM's PdBI at 2.7\,mm and an angular resolution of $\sim 1''$. Our main results and conclusions are summarised as follows:

\begin{enumerate}
\item Our continuum emission maps allow us to disentangle clearly the two protostellar objects, A1 and A2, present in IRAS\,4A. The envelope mass of A1 is likely a few times larger than that of A2.
\item We detected 23 relatively isolated lines from 11 different organic molecules including O-bearing COMs (e.g. CH$_3$CHO, HCOOCH$_3$, and CH$_3$OCH$_3$), N-bearing COMs (NH$_2$CHO, C$_2$H$_5$CN, and CH$_3$CN), and simpler species such as HNCO. The emission from all the detected molecules originates exclusively from A2, while A1 appears devoid of COM emission at the sensitivity of our observations even though it is a stronger continuum emitter.
\item At the wavelengths of our observations, we estimate that the optical depth of the dust emission is low in A2. On the other hand, dust emission may be optically thick in A1 at 1.2\,mm, but not sufficiently thick to hide completely the emission from a hypothetical hot corino in A1 that is similar in size and molecular contents to that present in A2. At 2.7\,mm, where dust emission is likely optically thin, the difference in the molecular emission between the two sources is still seen. Therefore, while longer wavelength observations of COMs are desirable to constrain their column densities in A1 better, the contrast found between the two protostellar sources in IRAS\,4A reflects a real difference in molecular richness.
\item In A2, we measured typical hot corino abundances with respect to H$_2$. These range between 10$^{-10}$ and $10^{-8}$ depending on the COM and are about an order of magnitude larger than those reported in previous, lower resolution studies of the same object. The deconvolved size of the hot corino in A2 is $\sim$\,70\,au.
\item For the first time, we were able to provide upper limits to the column densities and abundances of organic molecules in A1. The latter lie below 10$^{-10}$ for all species under optically thin dust emission conditions. A1 has lower molecular abundances, by about a factor 10, than those recently measured in the external gas envelope of prestellar cores or in the cold outer envelopes of other protostellar objects. One possible interpretation is that A1 is not yet sufficiently warm to have developed a hot corino and that its density is high enough for COMs and/or their precursors to be heavily frozen onto the mantles of dust grains in an environment that is well shielded from interstellar ultraviolet/cosmic ray radiation. This suggests that the inner regions of a young protostellar core that has not yet formed a hot corino may be poorer in COMs than the more external, lower density protostellar envelope. Alternatively, if a hot corino exists in A1, its size should be smaller than about 12\,au.
\item The most remarkable result of our study is the huge difference in COMs abundances existing between A1 and A2, which ranges between one and two orders of magnitude. Such a contrast may be due to protostar A2 being either more massive and/or subject to a higher accretion rate than A1, as a result of an inhomogeneous fragmentation of the parental molecular clump. This would naturally explain the smaller current envelope mass in A2 with respect to A1.
\item When comparing IRAS\,4A with I16293, another binary hot corino source, we notice that, in both objects, the weakest millimetre continuum source is the richest in molecular contents. Nevertheless, the two protostellar cores present in I16293 display hot corino activity, while only one of the two cores does in IRAS\,4A. This might be an indication that the latter is in an earlier evolutionary stage or that IRAS\,4A1 is undergoing a low-accretion phase that hinders strong hot corino activity.
\end{enumerate}

As a final note, we stress that the fact that chemical complexity may drastically vary within just a few tens or hundreds of au around protostars, in particular in multiple systems such as IRAS\,4A, highlights the importance of high angular resolution imaging to provide accurate constraints on the spatial distribution of molecules in these objects.

\begin{acknowledgements}
We are very grateful to C. Codella, C. Favre, C. Levefre, and H.B. Liu for useful discussions. We would also like to thank the GILDAS software team at IRAM for great help during data analysis. This paper makes use of the following ALMA data: ADS/JAO.ALMA\#2013.1.01102.S. ALMA is a partnership of ESO (representing its member states), NSF (USA) and NINS (Japan), together with NRC (Canada), NSC and ASIAA (Taiwan), and KASI (Republic of Korea), in cooperation with the Republic of Chile. The Joint ALMA Observatory is operated by ESO, AUI/NRAO and NAOJ. This paper makes use of data obtained with the IRAM PdBI. IRAM is supported by INSU/CNRS (France), MPG (Germany) and IGN (Spain). This study is supported by Grant-in-Aids from Ministry of Education, Culture, Sports, Science, and Technologies of Japan (21224002, 25400223, and 25108005). The authors acknowledge the financial support by JSPS and MAEE under the Japan-France integrated action programme (SAKURA: 25765VC).
\end{acknowledgements}

%
%

\bibliographystyle{aa} 

\bibliography{biblio} 



\end{document}